\documentclass[onefignum,onetabnum]{siamart171218}
\usepackage[margin=1in]{geometry}
\usepackage[round,numbers]{natbib}
\usepackage{setspace}
\usepackage{amsmath, mathtools}
\usepackage{mathrsfs}
\usepackage{graphicx}
\usepackage{stackengine}
\usepackage{caption}
\usepackage{subcaption}
\usepackage{epstopdf}
\usepackage{booktabs}
\usepackage{listings}

\usepackage{lipsum}
\usepackage{amsfonts}
\usepackage{graphicx}
\usepackage{epstopdf}
\usepackage{algorithm}
\usepackage{algorithmic}
\usepackage{xfrac}
\ifpdf
  \DeclareGraphicsExtensions{.eps,.pdf,.png,.jpg}
\else
  \DeclareGraphicsExtensions{.eps}
\fi

\newsiamremark{remark}{Remark}
\newsiamremark{hypothesis}{Hypothesis}
\crefname{hypothesis}{Hypothesis}{Hypotheses}
\newsiamthm{claim}{Claim}

\headers{Elliptic cross sections}{C.~Brimacombe, R.~M.~Corless, and M.~Zamir}

\title{Elliptic cross sections in blood flow regulation\thanks{Submitted to the editors DATE.
\funding{This work was supported by NSERC under grants numbered RGPIN-2020-06438 (RMC) and RGPIN-2019-04749 (MZ); and, while RMC was visiting the Isaac Newton Institute
during the programme
Complex Analysis: Tools, techniques, and applications, by EPSRC Grant \# EP/R014604/1.}}}

\author{Chris Brimacombe\thanks{University of Toronto, Toronto, \textsc{Canada}
  (\email{chris.brimacombe@mail.utoronto.ca }).}
\and Robert M.~Corless\thanks{Department of Computer Science, Western University and Cheriton School of Computer Science, University of Waterloo, London and Waterloo, \textsc{Canada}
  \email{rcorless@uwo.ca}.}
\and Mair Zamir\thanks{Department of Mathematics and Department of Medical Biophysics, Western University, London, \textsc{Canada}
  \email{zamir@uwo.ca}.}}

\usepackage{amsopn}

\newcommand{\Ce}{\ensuremath{\mathrm{Ce}}}
\newcommand{\ce}{\ensuremath{\mathrm{ce}}}

\newcommand{\se}{\ensuremath{\mathrm{se}}}

\newcommand{\eccentricity}{\ensuremath{\varepsilon}}
\newcommand{\fa}{\ensuremath{f_e}}
\newcommand{\ga}{\ensuremath{g_e}} 

\makeatletter
\newcommand*{\addFileDependency}[1]{
  \typeout{(#1)}
  \@addtofilelist{#1}
  \IfFileExists{#1}{}{\typeout{No file #1.}}
}
\makeatother

\newcommand*{\myexternaldocument}[1]{%
    \externaldocument{#1}%
    \addFileDependency{#1.tex}%
    \addFileDependency{#1.aux}%
}

\ifpdf
\hypersetup{
  pdftitle={Elliptic vascular cross sections in blood flow regulation},
  pdfauthor={C.~Brimacombe, R.~M.~Corless, and M.~Zamir}
}
\fi

\myexternaldocument{ex_supplement}

\renewcommand{\emph}[1]{\textsl{#1}}
\begin{document}
\maketitle

\begin{abstract}
{Arterial deformations arise in blood flow when surrounding tissue invades the space available for a blood vessel to maintain its circular cross section, the most immediate effects being a reduction in blood flow and redistribution of shear stress. Here we consider deformations from circular to elliptic cross sections. Solution of this problem in steady flow is fairly straightforward. The focus in the present paper is on pulsatile flow where the change from circular to elliptic cross sections is associated with a transition in the character of the equations governing the flow from Bessel to Mathieu equations. The study of this problem has been hampered in the past because of difficulties involved in the solution of the governing equations. In the present study we describe methods we have used to overcome some of these difficulties and present a comprehensive set of results based on these methods. In particular, vessel deformation is examined under two different conditions relevant to blood flow regulation: (i) keeping cross sectional area constant and (ii) keeping cross sectional circumference constant. The results provide an important context for the mechanism of neurovascular control of blood flow under the pathological conditions of vessel deformation.}
\end{abstract}
\textbf{Keywords:} Neurovascular control; Blood vessel deformation; Pulsatile blood flow; Coronary arteries; Mathieu equations/functions

\section{Introduction}
Arterial deformations arise in blood flow when surrounding tissue invades the space available for a blood vessel to maintain its circular cross section. This may occur in steady state when the invading tissue is pathological, or in oscillatory state when the invading tissue is driven by the effects of pulsatile blood flow. 

In the brain, the presence of a tumor may compress surrounding lymphatic and blood vessels, causing flow disruptions, especially within the restrictive environment of the rigid skull~\citep{tumor1,tumor2,tumor3}. In the heart, coronary vasculature embedded within the ventricular walls undergo periodic compression and deformation with each contraction of the heart muscle~\citep{zamir2006}. Segments of the aorta near the heart have also been reported to undergo periodic deformations from circular to elliptic cross section with each heart beat~\citep{moreno}. Coronary arteries tethered to the surface of the heart undergo a different kind of deformation as they are laterally displaced with each heart beat, causing a lateral acceleration of fluid and a lateral force on the tube wall, resulting in a change in its shape from a circular to an elliptic cross section~\citep{delfino}. Flow in tubes of noncircular cross sections, both steady and pulsatile, have also been discussed in relation to the movement of spinal fluids under normal and pathological conditions~\citep{irani,gupta,Kurtcuoglu,stockman,loth}.

It is well known that the flow in a tube of circular cross section is singular in the sense that any departure from the circular geometry of the cross section causes a reduction in the flow rate as well as a redistribution of the shear stress along the circumference of the tube wall whereby the shear stress at some points will be higher than that in an equivalent tube of circular cross section~\citep{haslam}. Both of these changes are important in blood flow, the latter in particular in relation to atherosclerosis~\citep{chatzizisis,malek,shaaban,himburg}. 

While blood vessel deformation by surrounding tissue may lead to many different forms of deformation of the vessel cross section, in the present study, to keep the problem mathematically tractable, we consider the limited problem of deformations from circular to elliptic cross sections. 

Flow within a blood vessel is generally under neurovascular control whereby a change in flow rate is mediated by a change of vessel diameter. The latter in turn is mediated by a change in muscular tension within the vessel wall to the effect of changing the length of the wall circumference~\citep{rowell}. If the vessel is deformed by surrounding tissue such that its cross section is transformed from circular to elliptic form, two distinctly different scenarios may follow, which we shall refer to as ``passive" and ``active" scenarios. Under a passive scenario the neurovascular control is absent, and a change from circular to elliptic cross section occurs with the circumference of the vessel wall remaining constant. Under the active scenario the neurovascular control responds by changing the tension within the vessel wall in an attempt to maintain the flow rate by keeping the cross sectional area available to the flow constant. The aim of the present study is to outline the analyses associated with these two scenarios and to present results illustrating the hemodynamic consequences in the two cases.

\begin{figure}
    \centering
    \includegraphics[width=0.5\textwidth]{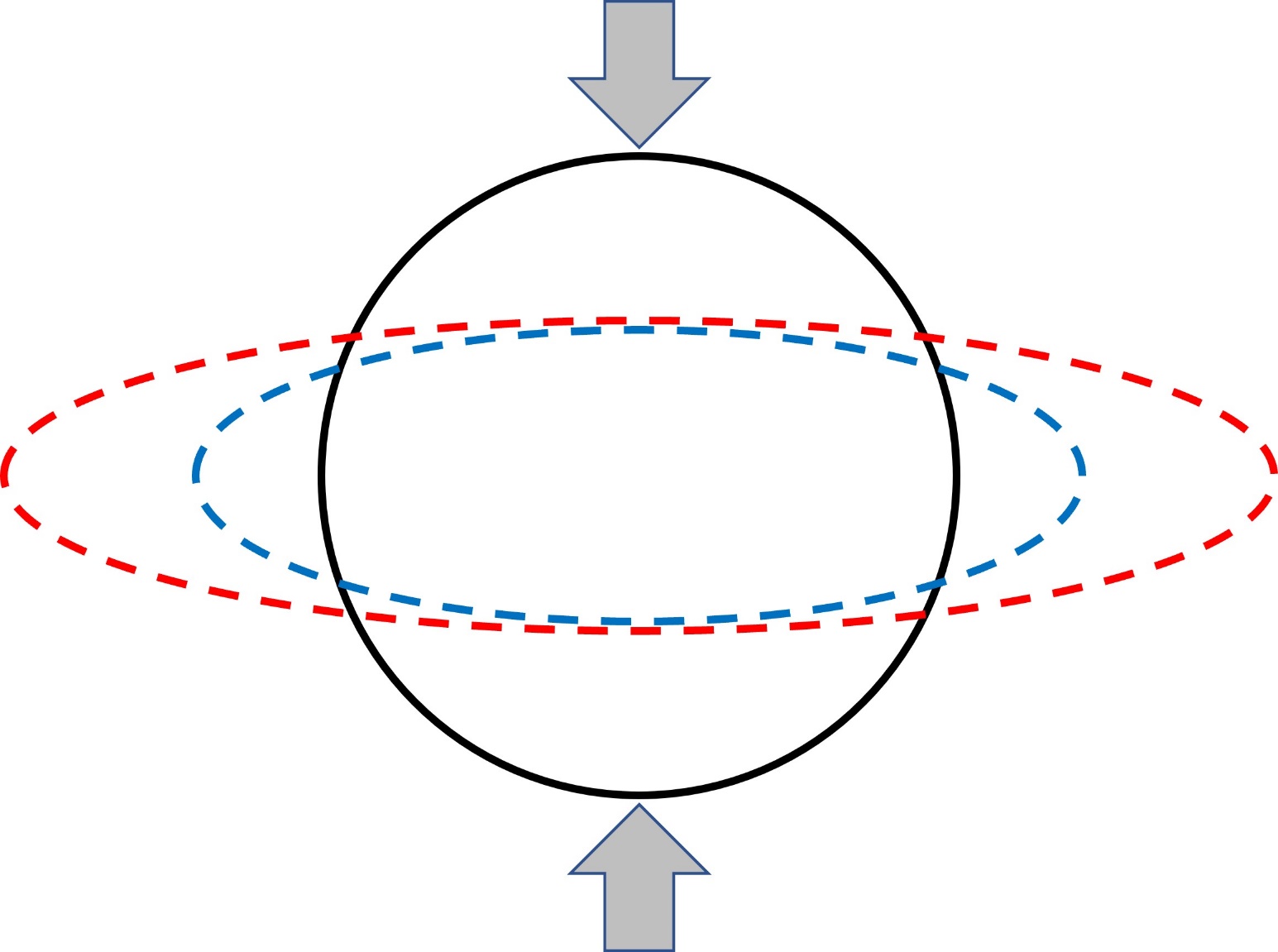}
    \caption{A blood vessel of circular cross section is compressed by surrounding tissue such that its cross section becomes elliptic with semiminor axis $\beta=\fa a$ where $\fa$ is a prescribed fraction of the circle radius $a$. Under a passive scenario (blue) regulatory control is absent and the length of circumference of the resulting ellipse is the same as that of the circle. Under an active scenario (red) the regulatory system intervenes in an attempt to keep the cross sectional area of the resulting ellipse the same as that of the circle. }
    \label{fig:illustration}
\end{figure}

While from a geometrical perspective the change from circular to elliptic cross sections may seem to be a ``smooth'' change, from a mathematical perspective it presents a discontinuity in the character of the equations governing pulsatile flow as well as in their solutions. Specifically, in the case of circular cross sections the equations governing the flow are Bessel equations and the solutions involve Bessel functions, while in the case of elliptic cross sections the flow is governed by Mathieu equations and the solutions involve Mathieu functions~\citep{haslam1998}. 

The study of pulsatile flow in tubes of elliptic cross sections has been hampered in the past because of difficulties involved in the solution of these equations and in the numerical evaluation of Mathieu functions with complex arguments~\citep{haslam,robertson,berselli,gupta,ziener}. In the present study we use a methodology described in~\cite{Brimacombe2021} to overcome these difficulties and to extend the range of ellipticity at which flow properties can be evaluated. In particular, the effects of vessel deformation on flow rate and on shear stress distribution along the vessel wall are presented. 
\section{Model equations and consequences}
\begin{figure}[t]
	\centering
    \includegraphics[width=0.7\textwidth]{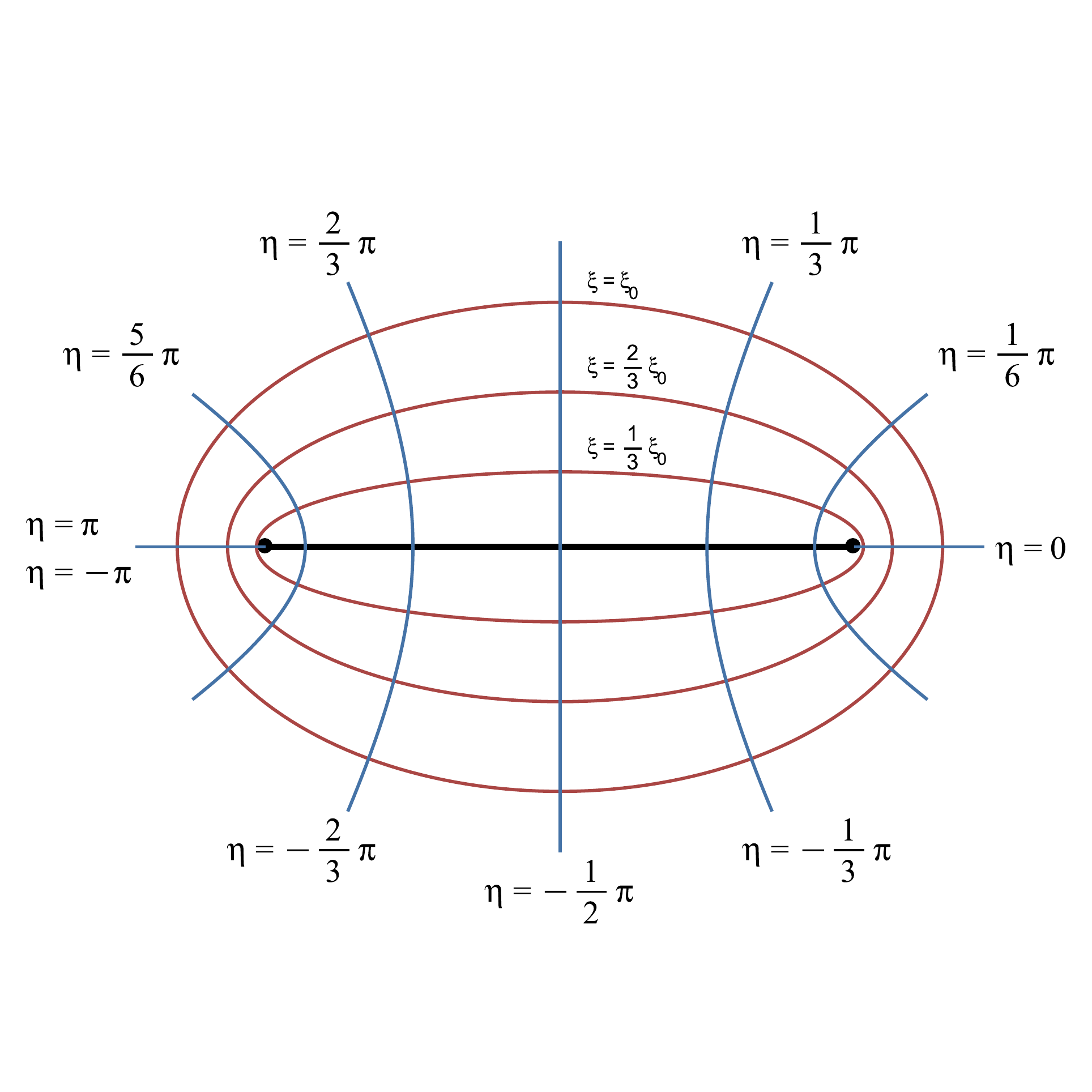}
    \caption{Confocal elliptic coordinate system $x=d\cosh\xi\cos\eta$, $y=d\sinh\xi\sin\eta$ used in the solution of the governing equations, where $\xi=\xi_0$ is the outer circumference of a cross section of the tube of elliptic cross section. Foci at $(\pm d, 0)$ indicated by solid black dots. The length of the semimajor axis of the outermost ellipse is $\alpha=d\cosh\xi_0$ and the length of the semiminor axis is $\beta=d\sinh\xi_0$.}
	\label{fig:EllipseAndNormalize}
\end{figure}

Consider an ellipse with semi-major and semi-minor axes, $\alpha$, $\beta$, respectively.
Figure~\ref{fig:EllipseAndNormalize} shows ellipses in confocal elliptic $\xi$, $\eta$ coordinates, which we will find useful.  If the foci are at $(\pm d, 0)$ then the normal Cartesian coordinates are $x = d\cosh\xi\cos\eta$ and $y=d\sinh\xi\sin\eta$.  If the parameter of the outer ellipse is $\xi_0$ then $\alpha = d\cosh\xi_0$ and $\beta = d\sinh\xi_0$. We have $d^2 + \beta^2 = \alpha^2$ from elementary geometry.
The eccentricity $\eccentricity$ of the outermost ellipse, at $\xi=\xi_0$, is defined by $\eccentricity = d/\alpha = \mathrm{sech}\xi_0$.  Thus the eccentricity of the confocal ellipses changes as $\xi_0$ changes.

Using polar coordinates $x=\alpha \sin \theta$ and $y=\beta \cos \theta$, the circumference of the ellipse is given by 
\begin{align}
&4 \int^{\pi/2}_{0} \sqrt{ \left( \frac{dx}{d\theta} \right)^{2} + \left( \frac{dy}{d\theta} \right)^{2} } \ d\theta \\
&= 4 \alpha \text{E}(\eccentricity)
 \label{eq:circumference}
\end{align}where
\begin{equation} \label{eq:I_{e}}
\text{E}(\eccentricity) = \int^{\pi/2}_{0} \sqrt{1 - \eccentricity^2 \sin^{2} \theta } \ d \theta
\end{equation}
is the complete elliptic integral of the second kind~\citep{lawden}.
\begin{figure}[t]
	\centering
    \includegraphics[width=0.6\textwidth]{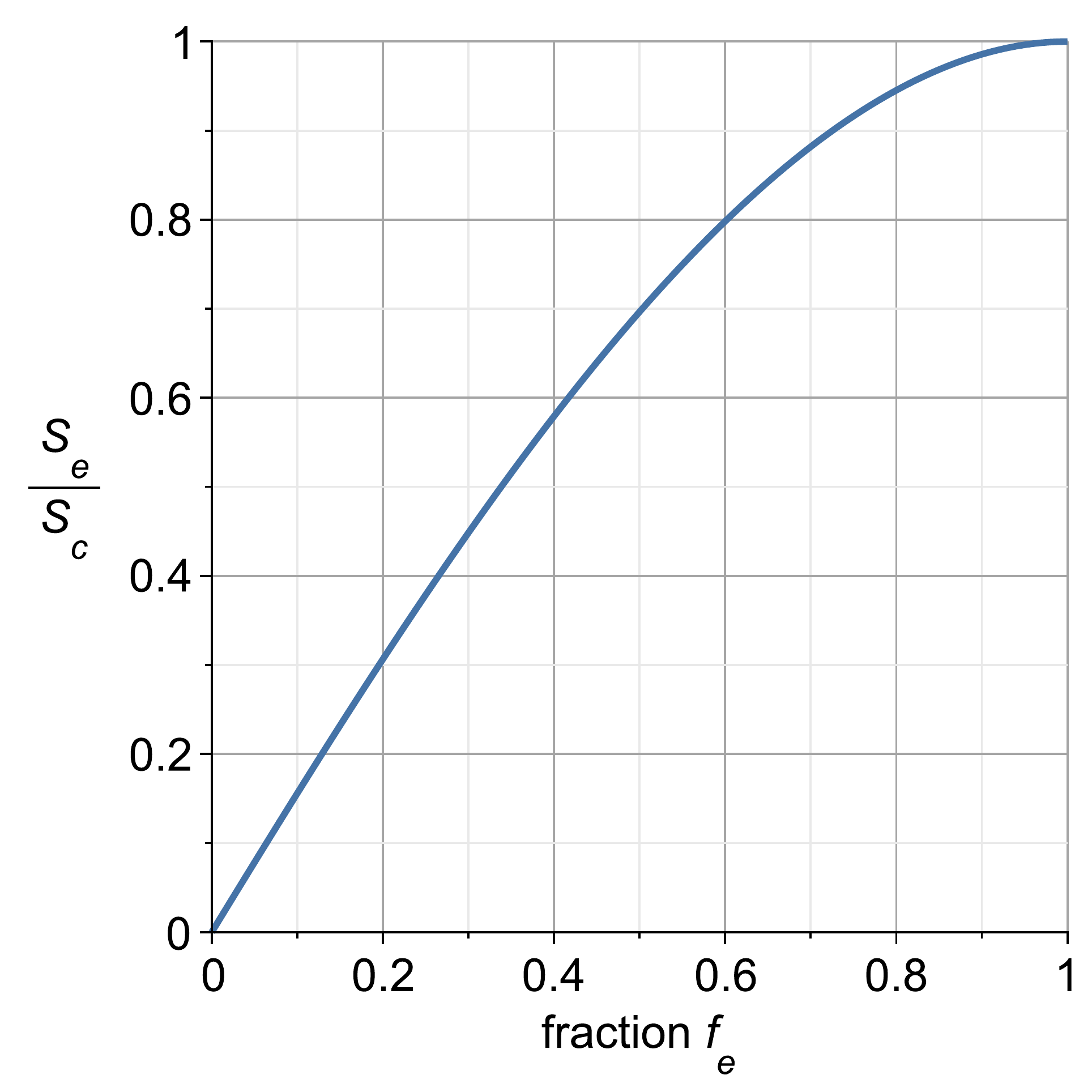}
    \caption{Relationships between areas of the circular and the elliptic cross sections when the length of their circumferences are the same.  $\text{S}_e$ and $\text{S}_c=$ are the areas of the elliptic and circular cross sections, respectively. In the limit, as the vessel cross section is flattened such that  $\fa \to 0$, the area of the ellipse vanishes.}
	\label{fig:allFigures}
\end{figure}
\paragraph{Passive scenario} Under this scenario the change from circular to elliptic cross section occurs while keeping the length of the circumference constant.

For an ellipse of eccentricity $\eccentricity$ and a circle of radius $a$ to have the same length of circumference, we have
\begin{equation}
2 \pi a = 4 \alpha \text{E}(\eccentricity) 
\end{equation} 
therefore 
\begin{align}
\frac{\alpha}{a} &= \frac{\pi}{2 \text{E}(\eccentricity)} \label{eq:majorAxisNormalized} \\ 
\frac{\beta}{a} &= \sqrt{1-\eccentricity^2} \frac{\pi}{2\text{E}(\eccentricity)}\label{eq:minorAxisNormalized}
\end{align}
If the area of the ellipse is denoted by $\text{S}_e$ ($=\pi\alpha\beta$) and the area of the circle is denoted by $\text{S}_c$ ($=\pi a^2$), then the ratio of the two is given by
\begin{equation}
\frac{\text{S}_{e}}{\text{S}_{c}} = \left( \frac{\pi}{2\text{E}(\eccentricity)}  \right)^{2} \sqrt{1-\eccentricity^2} \label{eq:areaOfEllipseNormalized}
\end{equation}
In the passive scenario, where the circumference remains constant on deformation from a circle of radius $a$, the foci of the ellipse are located at $(\pm d, 0)$ where
\begin{equation}
    d = \frac{\pi\eccentricity}{2E(\eccentricity)}\,a\>.
\end{equation}
This tends to $\pi a/2$ as $\eccentricity$ tends to $1$.

Compressing a circle of original radius $a$ to an ellipse with semi-minor axis $\beta = {\fa}a$ with ${\fa}<1$ while keeping the circumference $2\pi a = 4 E(\eccentricity)\alpha$ constant requires that $\alpha = {\ga} a$ where ${\ga} > 1$ is given by the following implicit formulae from equations~\eqref{eq:majorAxisNormalized}--\eqref{eq:minorAxisNormalized}:
\begin{align}
    {\fa} =& \frac{\pi \sqrt{1-\eccentricity^2}}{2E(\eccentricity)} \label{eq:transcendental} \\
    {\ga} =& \frac{\pi}{2E(\eccentricity)}\>.
\end{align}
To find ${\ga}$ for a given ${\fa}$ one must solve the transcendental equation~\eqref{eq:transcendental} for $\eccentricity$, and then use that in the equation for ${\ga}$. This is straightforward in Maple, by use of the command \texttt{fsolve}.  For convenience, we tabulate some fractions in Table~\ref{tab:fracs}.
\begin{table}
\centering
    \begin{tabular}{c|c|c}
        $\fa = \beta/a$ & $\eccentricity$ & $\ga = \alpha/a$ \\
    \midrule
        0.4 & 0.9611 & 1.448 \\
        0.5 & 0.9334 & 1.392 \\ 
        0.6 & 0.8925 & 1.330 \\
        0.7 & 0.8314 & 1.260 \\
        0.8 & 0.7359 & 1.182 \\ 
        0.9 & 0.5698 & 1.096 
    \end{tabular}
    \caption{Table of the major and minor axes, $\alpha$, $\beta$, of an ellipse having the same length of circumference of a circle of radius $a$, where $\eccentricity$ is the eccentricity of the ellipse.}
    \label{tab:fracs}
\end{table}

\paragraph{Active scenario} Under this scenario the change from circular to elliptic cross section occurs while keeping the cross sectional area constant.
In the active scenario, compressing a circle of radius $a$ so that its semi-minor axis $\beta = {\fa}a$ is a given fraction ${\fa}$ of the original radius, then since the area is $\pi\alpha\beta$ we must have $\alpha = a/{\fa}$.

\begin{figure}[t]
    \centering
    \includegraphics[width=0.6\textwidth]{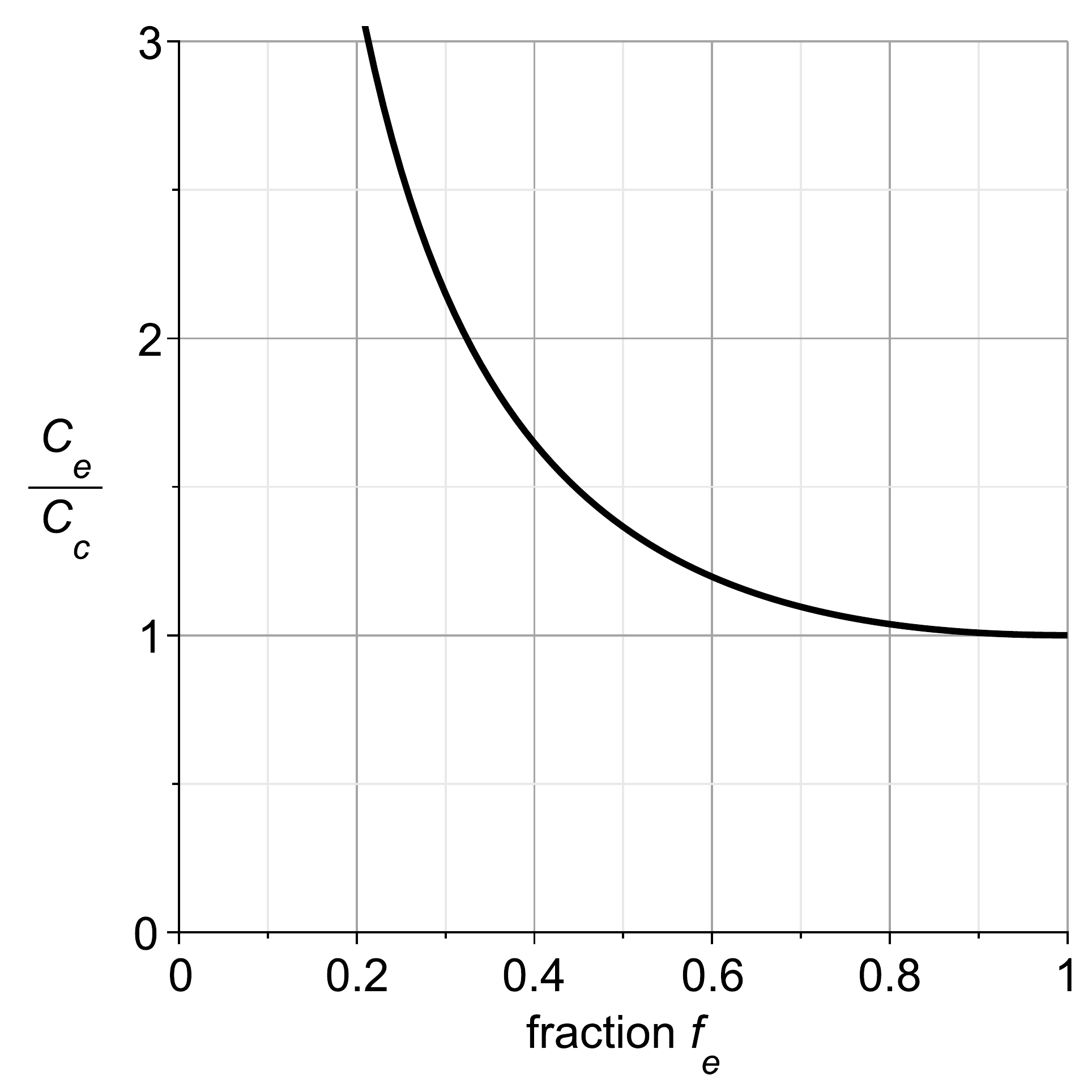}
    \caption{The ratio of the circumference of an ellipse to the circumference of a circle with the same area.  The formula is $C_e/C_c = 2E(\eccentricity)/(\pi(1-\eccentricity^2)^{1/4})$, with $E(\eccentricity)$ being the complete elliptic integral. As the fraction $f_e \to 0$, the eccentricity $\eccentricity \to 1$. The ratio of circumferences is singular as the fraction goes to $0$, as one would expect. The figure thus points to an intrinsic limitation of the regulatory system to increase the length of circumference of a blood vessel of elliptic cross section in an attempt to maintain its cross sectional area.
}
    \label{fig:ConstantAreaCircumferenceRatio}
\end{figure}

The ratio of the circumference of an ellipse to the circumference of a circle with the same area is
\begin{equation}
    \frac{C_{\mathrm{e}}}{C_{\mathrm{c}}} = \frac{2E(\eccentricity)}{\pi (1-\eccentricity^2)^{1/4}}\>.
\end{equation}
This is plotted in figure~\ref{fig:ConstantAreaCircumferenceRatio}.

In the active scenario, where  on deformation from a circle of radius $a$ the circumference is stretched by the regulatory system in order to keep the area constant, the foci of the ellipse are located at $(\pm d, 0)$ where
\begin{equation}
    d = \frac{\eccentricity}{(1-\eccentricity^2)^{1/4}}\,a\>.
\end{equation}
This is unbounded as $\eccentricity \to 1^{-}$.

\subsubsection*{Steady Flow in Tubes of Elliptic Cross Sections}
The properties of steady flow in a tube of elliptic cross section will be used as reference for the corresponding properties in pulsatile flow. The function governing the axial velocity $u_{0,e}$ is given by~\citep{zamir}:
\begin{equation}
u_{0,e} = -\frac{k_{0}\alpha^{2}\beta^{2}}{2\mu\left(\alpha^{2} + \beta^{2} \right)} \left(1 - \frac{x^2}{\alpha^{2}} - \frac{y^{2}}{\beta^{2}} \right)
\end{equation}
where $\mu$ is viscosity, and $k_{0}$ is the constant pressure gradient driving the flow. The maximum velocity occurs at $x=0,y=0$, the center of the ellipse: 
\begin{equation}\label{eq:steadyMaximumFlow}
\hat{u}_{0,e} = -\frac{k_{0} \alpha^{2}\beta^{2}}{2 \mu \left( \alpha^{2} + \beta^{2} \right)}
\end{equation} 
and volumetric flow rate is given by~\citep{zamir}
\begin{equation} \label{eq:qWithArea}
q_{0,e} = \frac{\hat{u}_{0,e} \text{S}_{e}}{2}
\end{equation}
Shear stress on the tube wall is given by~\citep{quadir}
\begin{equation}
\tau_{0,e}(x,y) = \frac{k_{0}\alpha^{2}\beta^{2}}{\alpha^{2}+\beta^{2}} \left( \frac{x^{2}}{\alpha^{4}} + \frac{y^{2}}{\beta^{4}} \right)^{1/2}
\end{equation}
Maximum shear occurs at the ends of the minor axis 
\begin{equation} \label{eq:steadMaximumShearStress}
\hat{\tau}_{0,e} = \frac{k_{0} \alpha^{2}\beta}{\alpha^{2} + \beta^{2}}
\end{equation}

Minimum shear occurs at the ends of the major axis
\begin{equation}
\Check{\tau}_{0,e} = \frac{k_{0}\alpha\beta^{2}}{\alpha^{2}+\beta^{2}}
\end{equation}


Maximum velocity and maximum shear are related by
\begin{equation} \label{eq:relationVelocityStress}
\hat{u}_{0,e} = - \frac{\beta}{2\mu} \hat{\tau}_{0,e}
\end{equation}

The corresponding quantities for a circle are obtained by letting $\beta \to \alpha$ or equivalently $\fa \to 1$. Then, for instance, the maximum and minimum shear both become $k_0 a/2$.

\begin{figure}
    \centering
    \includegraphics[width=0.6\textwidth]{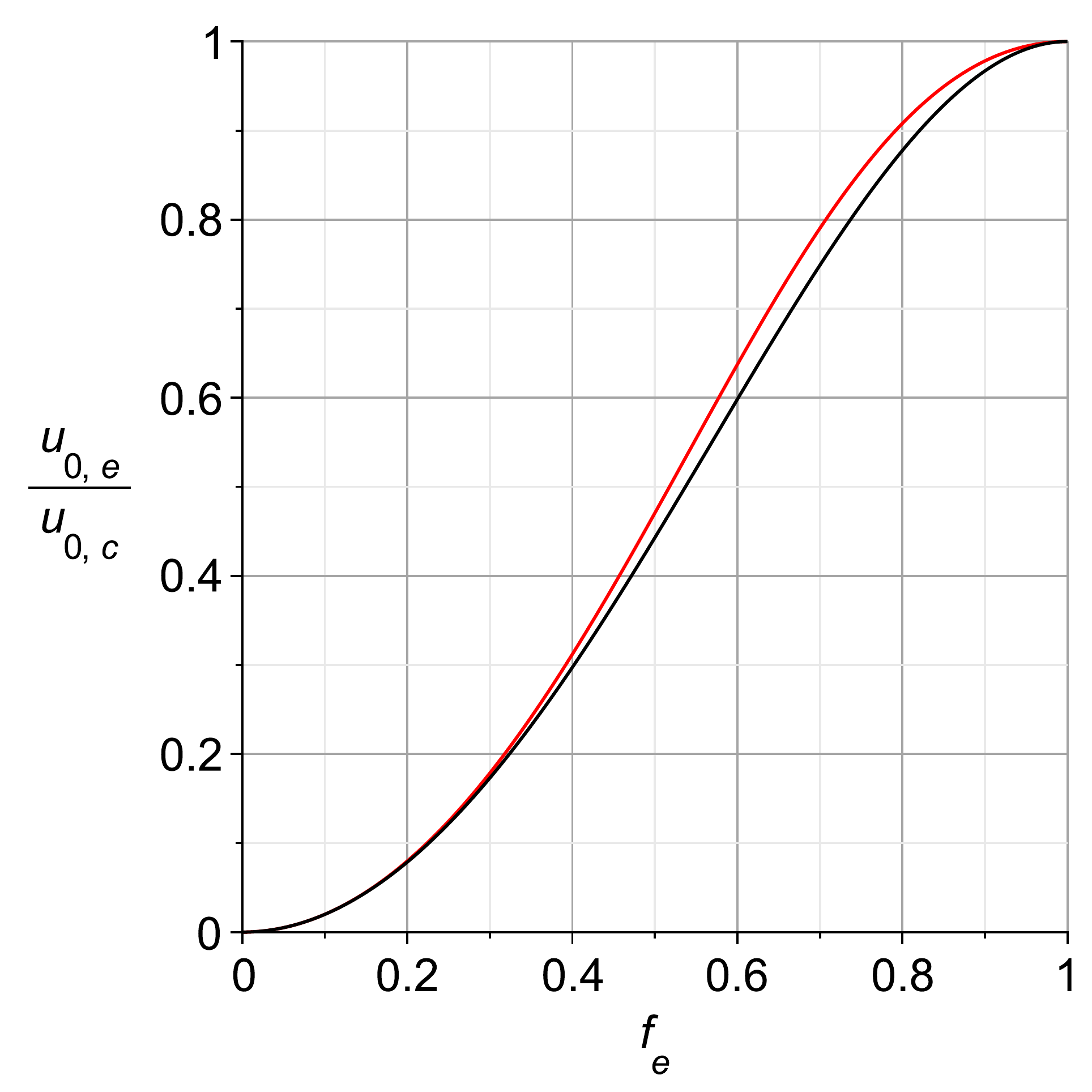}
    \caption{Maximum velocity in steady flow in a tube of elliptic cross section from equation~\eqref{eq:steadyMaximumFlow} compared to that in a tube of circular cross section in the two scenarios, active (red) and passive (black). }
    \label{fig:steadyvelocityactiveredpassiveblack}
\end{figure}

\begin{figure}
    \centering
    \includegraphics[width=0.6\textwidth]{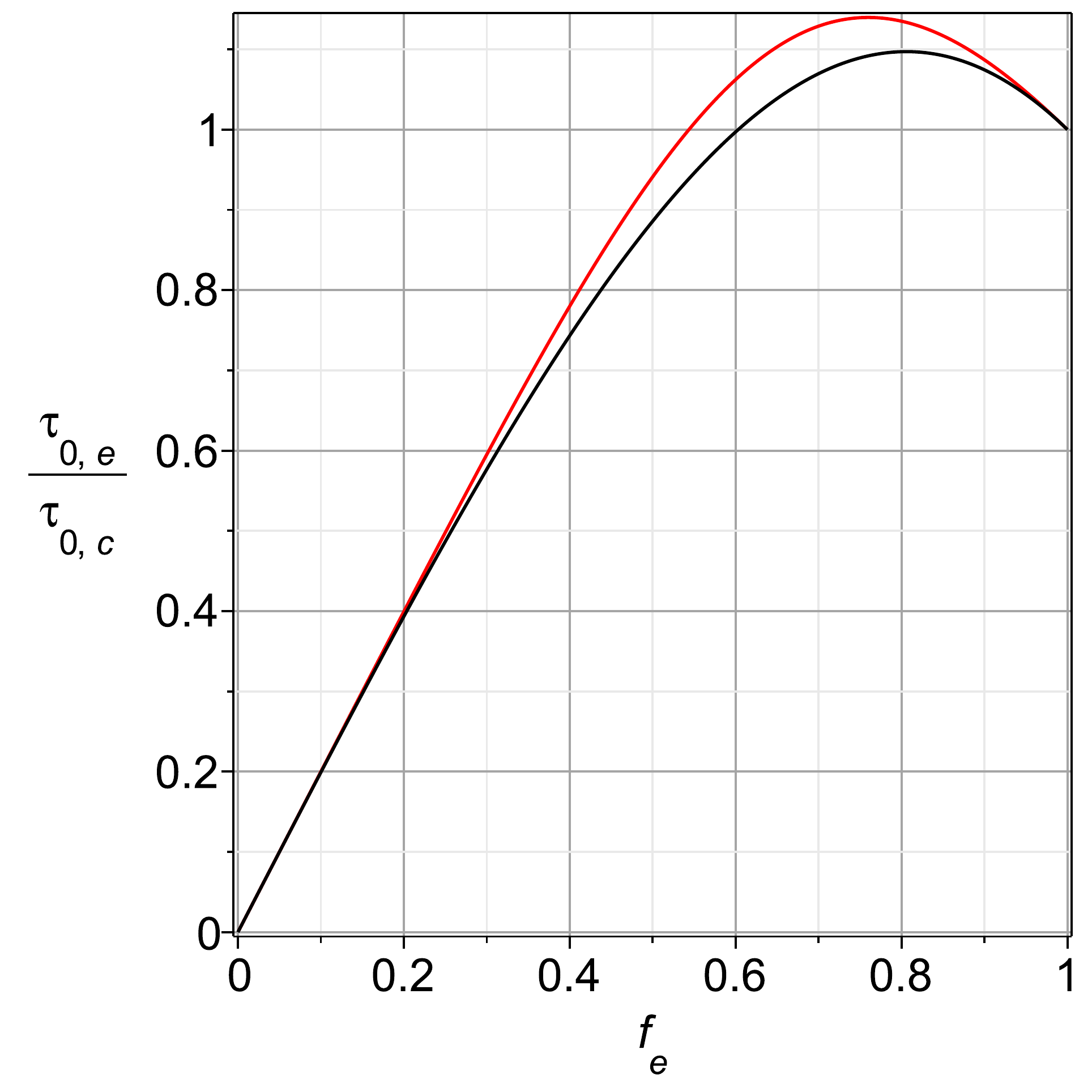}
    \caption{Maximum wall shear stress in steady flow in a tube of elliptic cross section from equation~\eqref{eq:steadMaximumShearStress} scaled by the constant shear stress on boundary of a tube of circular cross section and compared under the active (red) and passive (black) regulatory scenarios. }
    \label{fig:steadymaximumshearstress}
\end{figure}

\begin{figure}
    \centering
    \includegraphics[width=0.6\textwidth]{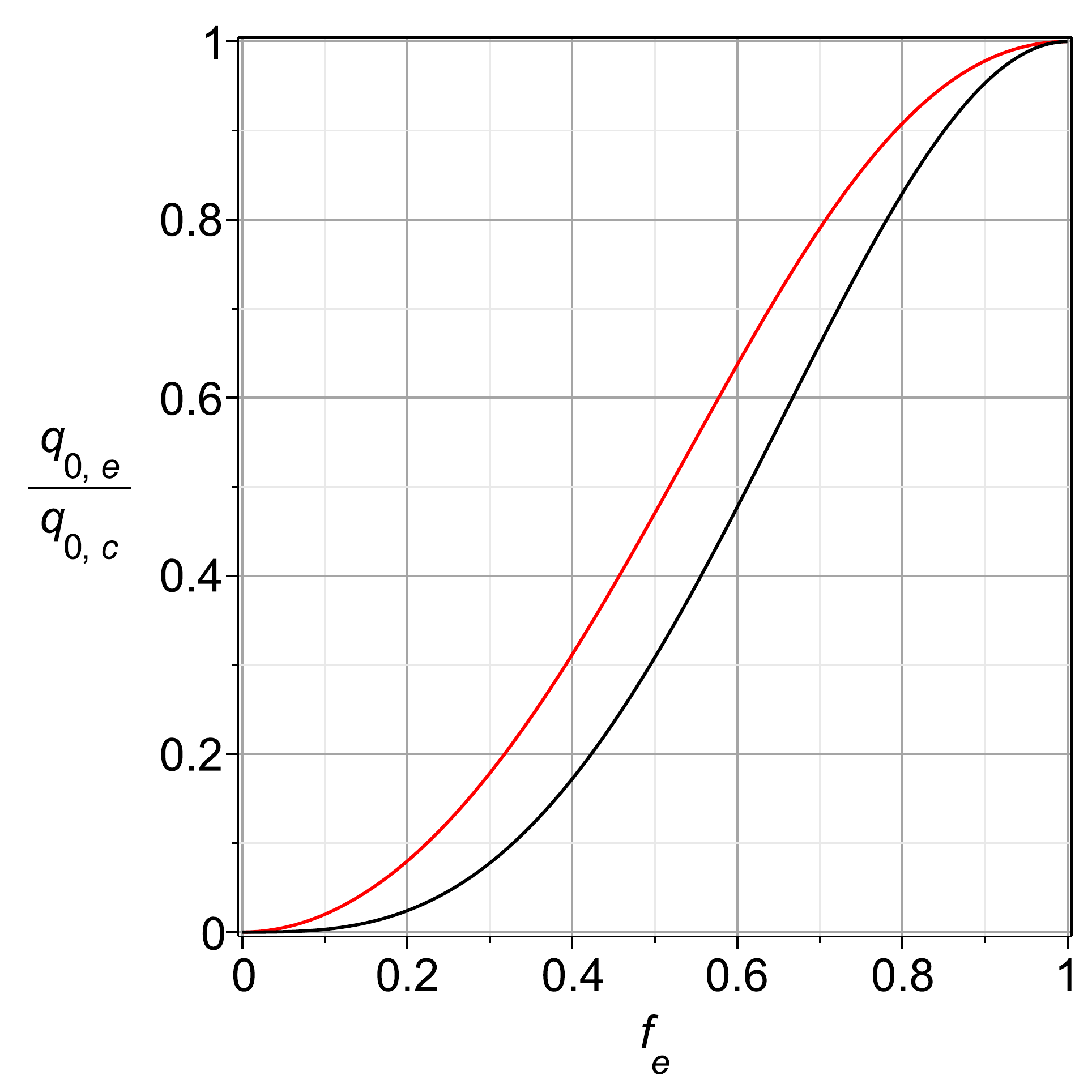}
    \caption{Flow rate in steady flow in a tube of elliptic cross section from equation~\eqref{eq:qWithArea} scaled by the flow rate in a tube of circular cross section and compared under the active (red) and passive (black) regulatory scenarios.}
    \label{fig:steadyflowrate}
\end{figure}

In Figures~\ref{fig:steadyvelocityactiveredpassiveblack}--\ref{fig:steadyflowrate} we use these formulas to compare the difference in \textsl{steady} flow in the two different scenarios, active (depicted with red curves in the figures) and passive (depicted with black curves).  All of the quantities above involve the semimajor and semiminor axes, $\alpha$ and $\beta$.  While for both scenarios $\beta = \fa a$ is the same, the value of $\alpha$ will be different in the active scenario ($a/\fa$) to the passive scenario $\ga a$ where $\ga$ is computed by solving a transcendental equation.  We see that there is indeed some difference in the flow quantities that arises in the two scenarios.


\subsubsection*{Pulsatile Flow in Tubes of Elliptic Cross Sections}
The axial velocity, $u_{e}$, in a tube of elliptic cross section can be written as the sum of a steady part, $u_{0,e}$, and an oscillatory part, $u_{\phi,e}$,
\begin{equation} \label{eq:originalU}
u_{e}(x,y,t) = u_{0,e} \left( x,y \right) + u_{\phi,e} \left( x,y,t \right)
\end{equation}
The equation governing the oscillatory part of the velocity is given by~\citep{haslam1998}
\begin{equation}\label{eq:originalGoverningEquation}
\frac{\partial u_{\phi,e}}{\partial t} + \frac{1}{\rho} \frac{\partial p}{\partial z} = \frac{\mu}{\rho} \left( \frac{\partial^{2}u_{\phi,e}}{\partial x^{2}} + \frac{\partial^{2}u_{\phi,e}}{\partial y^{2}} \right)
\end{equation}

The solution of this equation is facilitated by changing to the confocal elliptic coordinates shown in Figure~\ref{fig:EllipseAndNormalize}. {These were introduced by Lam\'e, who called them ``thermometric coordinates"\cite{mathieu1868memoire}.}
\begin{equation}
x = d \cosh \xi \cos \eta \>, \ y = d \sinh \xi \sin \eta
\end{equation} where the foci are at $(\pm d, 0)$ and $\xi$, $\eta$ are the elliptic coordinates. Using these confocal elliptic coordinates, an oscillatory pressure gradient of the form
\begin{equation} \label{eq:oscillatoryPressureGradient}
\frac{\partial p}{\partial z} = k_{0}e^{i\omega t}\>,
\end{equation}
and separation of variables
\begin{equation} \label{eq:uSeperation}
u_{\phi,e}(\xi,\eta,t) = w(\xi,\eta)e^{i\omega t}\>,
\end{equation} 
equation~\eqref{eq:originalGoverningEquation} can be formulated as an inhomogeneous Helmholtz equation
\begin{equation} \label{eq:governingEquationChanged}
\frac{2}{d^{2}(\cosh 2 \xi - \cos 2 \eta)} \left( \frac{\partial^{2}w}{\partial \xi^{2}} + \frac{\partial^{2}w}{\partial \eta^{2}} \right) - \frac{i \rho \omega}{\mu}w = \frac{k_{0}}{\mu}\>. 
\end{equation}
Using the translation
\begin{equation}
w(\xi,\eta) = v(\xi,\eta) - \frac{k_{0}}{i \rho \omega}\>, \label{eq:VandW}
\end{equation}
the inhomogeneous term of equation~\eqref{eq:governingEquationChanged} is eliminated and the equation becomes
\begin{equation} \label{eq:equationInV}
\left( \frac{\partial^{2} v }{\partial \xi^{2}} + \frac{\partial^{2} v}{\partial \eta^{2}} \right) - \frac{i}{2}\Lambda_{e} \left( \cosh 2 \xi - \cos 2 \eta \right) v = 0\>,
\end{equation}
where
\begin{equation}\label{eq:LambdaDef}
\Lambda_{e} = \frac{\rho \omega d^{2}}{\mu}
\end{equation}
is a nondimensional frequency parameter. 

The boundary conditions are given by
\begin{align}
v(\xi_{0},\eta) &= \frac{k_{0}}{i\rho \omega} \ (\text{no slip at tube wall}) \label{eq:noSlipCondition} \\
\frac{\partial v}{\partial \xi}\Bigr|_{\substack{\xi=0}} &= 0
\ (\text{symmetry})\\
v(\xi,0) &= v(\xi,\pi) \ (\pi \text{ periodic in } \eta)
\end{align}

While direct numerical solution of the governing equation equation~\eqref{eq:equationInV} is also possible, in this paper we pursue a solution based on the use of separation of variables, leading to the use of Mathieu functions. We do this in order to maintain the analytical connection with the classical solution of pulsatile flow in tubes of circular cross sections based on Bessel functions~\citep{zamir}. The use of Mathieu functions is not as straightforward as the use of Bessel functions, however, in part because of numerical difficulties in the evaluation of Mathieu functions of imaginary arguments.  Balancing that, this method is spectrally accurate, and does not require many eigenfunctions for the range of $q$ that we consider here.  Typically, we need only terms up to about $N=6$ or $N=8$.  
We postpone discussion of how to construct and evaluate the solution until section~\ref{sec:ComputeMathieu}.

In detail, the method proceeds as follows.  The treatment is standard, and we include it mostly for notation and readability.  Applying separation of variables to equation~\eqref{eq:equationInV} then yields two separate equations:
\begin{align}
\frac{d^{2} g}{d \eta^{2}} + \left( s - 2q \cos 2 \eta \right) g &= 0 \label{eq:ceMathieu} \\
\frac{d^{2} f}{d \xi^{2}} - \left( s - 2q \cosh 2 \xi \right) f &= 0 \label{eq:CeMathieu}
\end{align}where $s$ is a separating constant and
\begin{equation} \label{eq:qDefinition}
q = -\frac{i\Lambda_{e}}{4}\>.
\end{equation}
There is some risk of notational confusion because flow rate is often denoted by the variable $q$; here we will use flow variables with subscripts only, and the undecorated symbol $q$ will refer to the parameter in equation~\eqref{eq:qDefinition}.  This notation is standard for Mathieu functions, and we believe less confusion results when we use symbols in this fashion.

Equation~\eqref{eq:ceMathieu} is the Mathieu equation, and equation~\eqref{eq:CeMathieu} is the \textsl{modified} Mathieu equation.  These equations are equivalent, with the change of variable $\xi = i\eta$. The character of the solutions of the two equations are quite different, however.

Since in the present problem $\eta$ varies from 0 to $2\pi$, then $\eta$ must have periodicity $\pi$ or $2\pi$, which only occurs for discrete values of $s$, the eigenvalues of the Mathieu equation, designated by $s_m$ in~\citep{morse}. These eigenvalues are more commonly denoted nowadays with the letters $a_m$ and $b_m$; see the DLMF~\url{https://dlmf.nist.gov/28.2.ii}. This results in the set of Mathieu eigenfunctions $\ce_{m}$ and $\se_{m}$ for equation~\eqref{eq:ceMathieu}. Both $\ce_{m}$ and $\se_{m}$ are periodic, and $\ce_{m}$ is even, whereas $\se_{m}$ is odd. By convention, if $m$ is odd, then the Mathieu functions have period $2\pi$, while if $m$ is even, the Mathieu functions have period $\pi$.  In our problem, we are only interested in even $m$ values, so as to have $\pi$--periodicity, and even functions to satisfy symmetry along both axes\footnote{Symmetry breaking might very well be possible in a physical situation, and we believe it will be worthwhile to investigate this in future work}.

The even $\pi$-periodic solutions of Eq.\eqref{eq:ceMathieu} are the ordinary Mathieu functions denoted by $\ce_{2m}(\eta,q)$.  We will also need the modified Mathieu functions for the same $m$ and the same value of $q$, which are solutions of Eq.~\eqref{eq:CeMathieu}.  
The solution that we will compute will then be of the form
\begin{equation}
    v(\xi,\eta) = \frac{k_0}{i\rho\omega}\sum_{m\ge 0} b_{2m} \Ce_{2m}(\xi,q)\ce_{2m}(\eta,q)
\end{equation}
where the coefficients $b_{2m}$ will be determined by the no-slip boundary conditions, and we have taken the opportunity for a convenient scaling by $k_0/(i\rho\omega)$.

For certain values of $q$, however, such as the Mulholland--Goldstein value $q \approx 1.4688\,i $ (see~\cite{Brimacombe2021}), the Mathieu equation has \textsl{double eigenvalues} and at those points special care must be taken, because the ordinary Mathieu functions no longer form a complete set of orthogonal functions for expansion.  As $q$ tends to the Mulholland--Goldstein point, $\ce_0(\eta)$ and $\ce_2(\eta)$ coalesce and become the same function, and to ensure that expansion in these eigenfunctions is possible (also known as ``completeness''), a generalized eigenfunction must be added to the set of Mathieu functions.
In practice, as we will see, these isolated points make little difference to the solution because the overall problem is continuous (indeed analytic) in $q$, and so it is only the solution process which must be altered at these points.  Again, this is discussed in~\cite{Brimacombe2021}.

Using the no slip boundary condition from equation~\eqref{eq:noSlipCondition}, we find
\begin{equation} \label{eq:initialConditionForV}
v(\xi_{0},\eta) = \frac{k_{0}}{i \rho \omega} = \frac{k_{0}}{i \rho \omega}\sum_{m\ge 0} {b}_{2m} \Ce_{2m}(\xi_0,q)\ce_{2m}(\eta,q) \>.
\end{equation}
Since the Mathieu functions are orthogonal under the bilinear form
\begin{equation}
    \langle f, g \rangle := \int_{\eta=0}^{2\pi} f(\eta) g(\eta)\,d\eta
\end{equation}
(and if they have period $\pi$, the upper limit on the integral can be reduced to $\pi$), then multiplying equation~\eqref{eq:initialConditionForV} by $\ce_{2p}(\eta,q)$ and integrating with respect to $\eta$ gives
\begin{equation}
\int^{2 \pi}_{0} \ce_{2p}(\eta,q) \ d\eta = {b}_{2p}\Ce_{2p}(\xi_{0},q) \int^{2 \pi}_{0} \ce_{2p}^{2}(\eta,q) \ d\eta \>.
\end{equation}
We note that the bilinear form does not involve the complex conjugate.  Eigenvalues need not be real, and as parameters vary, eigenfunctions can coalesce.  Expansion in Mathieu functions is similar to harmonic expansion, but more complicated. In the usual case, when eigenvalues are simple, $b_{2m}$ is given by
\begin{equation} \label{eq:finalConstantC}
b_{2m} = \frac{\int^{2\pi}_{0} \ce_{2m}(\eta,q) \ d\eta}{\Ce_{2m}(\xi_{0},q)I_{2m}}\>.
\end{equation}
Here
\begin{equation} \label{eq:integralSimplify}
I_{2m} = \int^{2\pi}_{0} \ce_{2m}^2(\eta,q) \ d\eta\>.
\end{equation}
We compute these integrals by doing \textsl{exact} integration of the polynomial ``blends'' interpolating the solution, as described in section~\ref{sec:ComputeMathieu}. The computational cost for this is trivial.

\begin{remark}
The integral $I_{2m}$ can be zero.  In particular, if $q = 1.4688\ldots i$ (the Mulholland--Goldstein point mentioned earlier) then this integral is zero.  In this case, the expansion must be computed by a different method.  We ignore this possibility for the moment.
\end{remark}

\begin{remark}
    The value of $\Ce_{2m}(\xi_0,q)$ might \textsl{a priori} be zero.  In this case, we would have found a natural frequency of oscillation, and the solution would exhibit resonance.  We did not encounter resonance in any of the configurations we tried.  It seems that symmetric, even Mathieu functions with purely imaginary values of $q$ have no zeros on the imaginary axis, although we have not proved this.
\end{remark}

With this nonzero integral, 
equation~\eqref{eq:VandW} for $w$ becomes
\begin{equation} \label{eq:solutionOfW}
w(\xi,\eta) =  -\frac{k_{0}}{i\rho\omega} \left( 1 -  \sum_{m\ge 0} b_{2m} \Ce_{2m}(\xi,q)\ce_{2m}(\eta,q) \right)\>.
\end{equation}

\subsubsection*{Oscillatory Velocity}
The oscillatory flow velocity in a tube of elliptic cross section is then~\citep{haslam1998}
\begin{equation} \label{eq:velocityProfiles}
u_{\phi,e}(\xi,\eta,t) = \frac{4 \hat{u}_{0,e}}{i \lambda_{e}} \left( 1 -  \sum_{m\ge 0} b_{2m} \Ce_{2m}(\xi,q)\ce_{2m}(\eta,q) \right) e^{i \omega t}
\end{equation}
where
\begin{equation}
    \lambda_e = \frac{1}{2}\sinh2\xi_0\tanh2\xi_0 \Lambda_e = \frac{2(1-\eccentricity^2)}{\eccentricity^2\left(2-\eccentricity^2\right)}\Lambda_e 
\end{equation}
is a second nondimensional frequency parameter.

\subsubsection*{Oscillatory Flow Rate}
The flow rate is obtained by integrating the oscillatory velocity over the elliptic cross section 
\begin{align} 
q_{\phi,e}(t) &= \iint\limits_D u_{\phi,e}(\xi,\eta,t) \ dA \\
&= \left( \iint\limits_D v(\xi,\eta) \ dA + \frac{4\hat{u}_{0,e}}{i \lambda_{e}} \iint\limits_{D} \ dA \right) e^{i \omega t} \label{eq:beginningQ}
\end{align}
where $D$ is the region enclosed by the bounding ellipse. The second integral on the right side of equation~\eqref{eq:beginningQ} can be evaluated analytically:
\begin{equation} \label{eq:rightHandIntegral}
\frac{4 \hat{u}_{0,e}}{i \lambda_{e}} \iint\limits_D \ dA = \frac{8 q_{0,e}}{i \lambda_{e}}
\end{equation}
where $q_{0,e}$ is the steady flow rate in a tube of elliptic cross section (equation~\eqref{eq:qWithArea}). The first integral on the right hand side of equation~\eqref{eq:beginningQ} is then evaluated using $v$ in equation~\eqref{eq:equationInV} 
\begin{equation} \label{eq:delSquaredIntegral}
\iint\limits_D v \left( \xi, \eta \right) \ dA = \frac{\mu}{i\rho\omega} \iint\limits_D \frac{2}{d^{2}(\cosh 2 \xi - \cos 2 \eta)} \left( \frac{\partial^{2}v}{\partial \xi^{2}} + \frac{\partial^{2}v}{\partial \eta^{2}} \right) \ dA
\end{equation}
If $n$ is an outward pointing normal and $ds$ is an elemental surface, then by Green's theorem it follows that
\begin{equation} \label{eq:42}
\frac{\mu}{i\rho\omega} \iint\limits_D \frac{2}{d^{2}(\cosh 2 \xi - \cos 2 \eta)} \left( \frac{\partial^{2}v}{\partial \xi^{2}} + \frac{\partial^{2}v}{\partial \eta^{2}} \right) \ dA = \frac{\mu}{i\rho\omega} \int_{\partial D} \frac{\partial v}{\partial n} \ ds
\end{equation}
where $\partial D$ is the positively oriented bounding curve of $D$. 

It is shown in McLachlan~\citep{mclachlan} that $ds = \delta d \eta$ and $dn = \delta d \xi$, where
\begin{equation} \label{eq:delta}
\delta = d \left( \cosh^{2} \xi - \cos^{2} \eta \right)^{1/2}
\end{equation}
Thus, equation~\eqref{eq:42} becomes
\begin{equation} \label{eq:surfaceArcToEta}
\frac{\mu}{i\rho\omega} \int_{\partial D} \frac{\partial v}{\partial n} \ ds = \frac{\mu}{i\rho\omega} \int^{2\pi}_{0} \left( \frac{\partial v}{\partial \xi} \right)_{\xi = \xi_{0}} d \eta
\end{equation}
and 
\begin{equation}
   \left( \frac{\partial v}{\partial \xi} \right)_{\xi = \xi_{0}} =  \frac{k_{0}}{i \rho \omega}  \sum_{m\ge 0} {b}_{2m} \Ce_{2m}'(\xi_0,q)\ce_{2m}(\eta,q)  \label{eq:normalDerivative}
\end{equation}
where $'$ here denotes differentiation with respect to $\xi$. Because blends are polynomials, differentiation with them is simple, and the code we use provides for this automatically\footnote{It is important to remember that because the code implements $\Ce_{2m}(\xi,q)$ by $\ce_{2m}(i\xi,q)$ one has to use the chain rule and multiply by $i$: $\Ce_{2m}'(\xi,q) = i\ce_{2m}'(i\xi,q)$.}.  We thus get (apart from rounding errors) exact derivatives of the interpolants being used to represent the solutions.  Because the solutions are so high-order, the derivatives are themselves accurate: while they typically lose an order of accuracy for each derivative taken, if one starts with order $16$ then taking one derivative does not do much harm.  We remark that with high enough frequency, however, which does occur with large eigenvalues for Mathieu functions, one would need to work to higher precision to maintain this accuracy.  For the computations of this paper, we only used higher precision to check the numerics, and found double precision to be perfectly satisfactory.

Integration of this formula with respect to $\eta$ is straightforward, using the exact quadrature formula for blendstrings.  But in fact we have already integrated each of these functions, in computing the $b_{2m}$.  NB: if the integrals were only to $\pi$ and not to $2\pi$, one must multiply the following formula by $2$.

 Using equation~\eqref{eq:integralSimplify} 
we find that
\begin{equation}
 \frac{\mu}{i\rho\omega} \int^{2\pi}_{0} \left( \frac{\partial v}{\partial \xi} \right)_{\xi = \xi_{0}} d \eta = -\frac{\mu k_0}{ \rho^2 \omega^2} \sum_{m\ge 0} b_{2m}^2 I_{2m} \Ce_{2m}'(\xi_0,q)\Ce_{2m}(\xi_0,q)\>.    
\end{equation}
If we further use the relation
 \begin{equation}
 \frac{\lambda_e\mu}{\rho\omega\text{S}_{e}} = \frac{1}{\pi} \tanh 2 \xi_{0}\>,
 \end{equation}
this implies that the oscillatory flow rate in a tube of elliptic cross section is given by the following (cf.~\citep{haslam1998}):
\begin{equation} \label{eq:oscillatoryVolumetricFlow}
q_{\phi,e}(t) = \frac{8q_{0,e}}{i \lambda_{e}} \left( 1 - \frac{1 }{i \pi \lambda_{e}} \tanh 2 \xi_{0} \sum_{m\ge 0} b_{2m}^2 I_{2m} \Ce_{2m}'(\xi_0,q)\Ce_{2m}(\xi_0,q) \right)e^{i \omega t} \>.
\end{equation}
\subsubsection*{Oscillatory Wall Shear Stress}

By its definition, the wall shear stress is given by 
\begin{equation} \label{eq:partsOfTau}
\tau_{\phi,e} \left(\eta,t \right) = \mu \left( \frac{\partial u_{\phi,e}}{\partial n} \right)_{\partial D} = \mu \left( \frac{\partial v}{\partial n} \right)_{\partial D}e^{i \omega t} 
\end{equation}
where $u_{\phi,e}$ is the oscillatory velocity in a tube of elliptic cross section (equation~\eqref{eq:velocityProfiles}). Using the elemental arc length analysis of McLachlan~\citep{mclachlan}, it can be shown that
\begin{equation}\label{eq:partsOFellipticArc}
\left( \frac{\partial v}{\partial n} \right)_{\partial D} = \frac{1}{\delta_0} \left( \frac{\partial v}{\partial \xi} \right)_{\xi = \xi_{0}}
\end{equation}
where $\delta_0$ is 
\begin{equation} \label{eq:delta0}
\delta_0 = d \left( \cosh^{2} \xi_0 - \cos^{2} \eta \right)^{1/2}
\end{equation}
Substituting from equation~\eqref{eq:normalDerivative} for the derivative on the right hand side, this becomes
\begin{equation}\label{eq:secondsAway}
\left( \frac{\partial v}{\partial n} \right)_{\partial D} = \frac{1}{\delta_0}\left(\frac{k_{0}}{i \rho \omega}  \sum_{m\ge 0} {b}_{2m} \Ce_{2m}'(\xi_0,q)\ce_{2m}(\eta,q) \right)\>.
\end{equation}
Using equation~\eqref{eq:steadMaximumShearStress} we can replace $k_0$ by $\hat\tau_{0,e}(\alpha^2+\beta^2)/(\alpha^2\beta)$, or more conveniently by the limiting case of the circle: $k_0 = 2\hat\tau_{0,c}/a$.  Remember that $a$ is the radius of the original circle, and $\beta = \fa a$.
After some algebra we obtain the following expression for oscillatory wall shear stress in a tube of elliptic cross section:
\begin{equation} \label{eq:oscillatoryShearStress}
\tau_{\phi,e}\left( \eta, t \right) =
\frac{ 4\beta\fa \hat{\tau}_{0,c}  }{i\delta_0 \lambda_{e}(2-\eccentricity^2)} \left( \sum_{m\ge 0} {b}_{2m} \Ce_{2m}'(\xi_0,q)\ce_{2m}(\eta,q)\right)e^{i \omega t} 
\end{equation}

For reference and comparison, the (constant) oscillatory wall shear stress in a tube of circular cross section is given by the following~\cite{zamir2000}:
\begin{equation}
    \tau_{\phi,c} = \frac{2\tau_{0,c}}{\Lambda_c} \frac{J_1(\Lambda_c)}{J_0(\Lambda_c)} \label{eq:circlewallshearstress}
\end{equation}
where $J_k(z)$ for $k=0$, $1$ are Bessel functions of the first kind, $\tau_{0,c} = k_0 a/2$,  and
\begin{equation}
    \Lambda_c = \left(\frac{i-1}{\sqrt2}\right)\sqrt{\frac{\rho \omega}{\mu}}\,a\>.
\end{equation}
\section{Computation with Mathieu functions\label{sec:ComputeMathieu}}

We will not review all existing numerical methods for computing with Mathieu functions here, but instead refer to~\cite{Brimacombe2021}, which is available as an open-access article.
We will, however, summarize the method that we actually used, and give a few more details about the method in a subsection that may be skipped by a reader more concerned with the results, as opposed to how we got them.

For notational convenience we refer to values of $q$ with positive imaginary part, but because the eigenvalues are the same for $q$ and $-q$ in the even and symmetric case (see eg.\ the DLMF \url{https://dlmf.nist.gov/28.2}), this is sufficient for our application (which has negative imaginary part) and saves writing many minus signs.

The previous work of Haslam and Zamir in~\citep{haslam} used truncations of an infinite tri-diagonal eigenvalue-eigenvector problem to obtain approximations to the eigenvalues $a_{2m}$. This method goes back at least to the work of Ince, and is widely used~\cite{Brimacombe2021}.  The matrix in question, for the even and symmetric eigenfunctions, is

\begin{equation}
 \left[ \begin {array}{cccccc} 0&\sqrt {2}q&0&0&0&\cdots\\ \noalign{\medskip}
\sqrt {2}q&4&q&0&0&\cdots\\ \noalign{\medskip}0&q&16&q&0&\cdots\\ \noalign{\medskip}0
&0&q&36&q &\cdots\\
\noalign{\medskip}0
&0&0&q&64 &\ddots\\
\noalign{\medskip}\vdots&\vdots&\vdots&\vdots&\ddots&\ddots\end {array} \right]
\left[ \begin{array}{c}
\sqrt{2}A_0\\
\noalign{\medskip}
A_2\\
\noalign{\medskip}
A_4\\
\noalign{\medskip}
A_6\\
\noalign{\medskip}
A_8\\
\noalign{\medskip}
\vdots
\end{array}
\right]
= \lambda\left[ \begin{array}{c}
\sqrt{2}A_0\\
\noalign{\medskip}
A_2\\
\noalign{\medskip}
A_4\\
\noalign{\medskip}
A_6\\
\noalign{\medskip}
A_8\\
\noalign{\medskip}
\vdots
\end{array}
\right]\>.
\label{eq:tridceeven}
\end{equation}
Truncation at ``large enough" dimension gives good estimates of the eigenvalues, but there is a question of exactly how large should we take the matrix, and once the eigenvalues have been computed, how accurate they are.  Notice that this is a \textsl{complex symmetric} matrix, not a Hermitian matrix.

In our computations, we start with the matrix method, but only to get initial estimates of the eigenvalues $a_{2m}$.  We then apply the continued fraction method of Blanch as described in~\cite{Brimacombe2021} and use Newton's method to refine the eigenvalues to the desired accuracy.  This tells us precisely how accurate each eigenvalue is, and is more efficient than computing larger and larger matrices until the eigenvalues converge.  Our procedure works well enough for all simple eigenvalues, although sometimes we have to increase precision.  For the double eigenvalues, we proceed differently.

Double eigenvalues occur for purely imaginary $q$, but (as elsewhere in the complex $q$-plane) only at isolated points: At the Mulholland--Goldstein point $q \approx 1.4688 i$, and (next smallest) $q \approx 16.47 i$, and so on.  We have pre-computed several of these by the method of Hunter and Guerrieri~\cite{hunter1981eigenvalues}. They are tabulated in~\cite{Brimacombe2021} and are also available on-line in the code repository for that paper.

Given numerical values for the semimajor axis $\alpha$ and semiminor axis $\beta$, and given a numerical value for the (purely imaginary) parameter $q$, we computed up to certain index $N$ (frequently $N$ was $6$ and sometimes $8$; because this is a spectral method, convergence is very rapid) of the Mathieu eigenvalues $a_{2m}(q)$, for $m=0$, $1$, $\ldots$, $N$.  If the eigenvalues were distinct (which was usually, but not always, the case) then we computed the Mathieu functions $\ce_{2m}(\eta)$ on the interval $0 \le \eta \le \pi$ and the corresponding modified Mathieu functions $\Ce_{2m}(\xi)$ on the interval $0 \le \xi \le \xi_0 = \mathrm{invcosh}(\alpha/d) = \mathrm{invsech}(d/\alpha) = \mathrm{invsech}(\eccentricity)$. This is because $\alpha = d \cosh \xi_0$ or, more simply, $\eccentricity = \mathrm{sech}\xi_0$ gives the value $\xi_0$ of the parameter $\xi$ at the tube wall\footnote{Here, we are using David Jeffrey's notation for functional inverses: $y = \mathrm{invcosh}(x)$ means $x = \cosh(y)$, etc. This notation is superior for branched inverses, and superior pedagogically even for simple functions, to the more common overloading of superscripts or use of the inappropriate word ``arc", and we hope that it catches on. }.  

To compute the Mathieu functions and modified Mathieu functions, we used the Hermite-Obreshkov integrator sketched in~\cite{Brimacombe2021}.  We worked in double precision (except where noted explicitly here) and typically used an order $30$ or $40$ method, with grade\footnote{The word ``grade'' means ``degree at most''.  This is convenient because the final Taylor coefficients computed might be zero, but this is still useful information.} $15$ or $20$ Taylor series computed on each marching step, and ``blendstrings'' as piecewise polynomial interpolants giving the value of the solution (and whatever derivatives were required).

\subsection{More details of the numerical method\label{sec:details}}
We treat the Mathieu equation (and the modified Mathieu equation) as an initial-value problem (IVP) for an ordinary differential equation (ODE), once both $q$ and the eigenvalue $s$ are fixed.  To compute the Mathieu function, we could use almost any standard method to solve the IVP\footnote{We reassure the reader that we do know and highly value the standard general methods, as described for instance in the classic~\cite{Hairer(2002),Hairer(1993)}. We are also aware of the truly remarkable advances made since then, such as are described in~\cite{Rackauckas2017}.  We have even contributed to the literature and the software ecosystem in the past~\cite{Shampine2000}. But while writing a special-purpose solver for the Mathieu equation---when so many good solvers already exist---might seem quixotic, bear with us for a bit: it turns out to be useful and we believe interesting, and in particular it is reassuring to have the ability to retrospectively measure how accurate the solutions are. }. But the modified Mathieu equation is related to the Mathieu equation by the change of variables $\xi = i\eta$.  That is, if the standard method chosen for the Mathieu equation could work in the complex plane, then it could also be used for the modified Mathieu equation.  This idea restricts us to implementations that work over the complex plane, but because we have a complex parameter $q$ (in fact, purely imaginary in our application), this is necessary anyway.

Also, there is an opportunity for greater efficiency and control. Since the Mathieu equation is linear, special-purpose methods appropriate for linear problems might be used.  More, since the Mathieu equation can be written in a ``D-finite" or ``holonomic" form\footnote{This fact was already known to Mathieu, although the names $D$-finite or holonomic had not been invented yet in 1868.  But writing the differential equation in this form allows for faster human computation, too.}, Taylor series coefficients can be computed rapidly given the initial values $y(\eta_n)$ and $y'(\eta_n)$.  In fact, we do not use the D-finite form even though it does offer the potential of significant speed-up~\cite{mezzarobba2010numgfun}; this might be pursued in future.  Straightforward generation of Taylor coefficients by Cauchy convolution with those of $\cos 2\eta$ was fast enough for our purposes.

\subsubsection{Blends\label{sec:blends}}
We now explain the interpolants that we use.
``Blends'', or two-point Hermite interpolants, are described in~\cite{corless2020blends}.
In brief, if one knows Taylor coefficients $p_j$ for $0 \le j \le m$ at one end $z_k$ of an interval, and Taylor coefficients $q_j$ for $0 \le j \le n$ at the other end $z_{k+1}$ of an interval, and $z = z_k + sh$ where $h = z_{k+1}-z_k$ is the width of the interval so that $0 \le s \le 1$, then the following polynomial ``blends" the two sets of Taylor coefficients together to form an excellent approximation of the function over the interval:
(Hermite, Cours d'Analyse 1873)
\begin{align}
    H_{m,n} \! \left(s \right) = &
\sum_{j=0}^m \sum_{k=0}^{m-j} \binom{n+k}{k} s^{k +j} \left(1-s \right)^{n +1}  p_{j} \nonumber\\
& + \sum_{j=0}^n \left(-1\right)^{j} \sum_{k=0}^{n-j} \binom{m+k}{k} s^{m +1} \left(1-s \right)^{k +j} q_{j}
\end{align}
has $H^{(j)}(0)/j! = p_j$ for $0 \le j \le m$ and $H^{j}(1)/j! = q_j$ for $0 \le j \le n$. In this formula, differentiation is with respect to $s$, and care must be taken to include the correct factors of $h$ from the chain rule when using the formula for the interval $[z_k,z_{k+1}]$.

The error in Hermite interpolation is known; the results on the real line are given in~\cite{Kansy(1973)} (and the complex results were known to Hermite).  Here, the general real results simplify to
\begin{equation}
f(s) - H_{m,n}(s) = \frac{f^{(m+n+2)}(\theta)}{(m+n+2)!} s^{m+1}(s-1)^{n+1} \label{eq:errorblend}
\end{equation}
for some $\theta = \theta(s)$ between $0$ and $1$. 

If we have a sequence of nodes, say $z_k$ for $0 \le k \le M$, where Taylor coefficients for an analytic function $f(z)$ are known up to grade (say) $m_k$ at each node, then it is natural to approximate $f(z)$ on each segment from $z=z_k$ to $z=z_{k+1}$ by the blend determined by those two sets of Taylor coefficients. This gives a piecewise polynomial interpolant, which we call a  ``blendstring'' for short.  

For instance, if Taylor series of only grade $1$ are used at each node, then the blendstring is just the more familiar pure piecewise cubic Hermite interpolant on each subinterval, and the result is similar to a cubic spline.  Taylor series of only grade $1$ do not give us the needed accuracy, though, and we always use much higher order.   

As described in~\cite{corless2020blends}, these interpolants are remarkably stable numerically, even for ludicrously high order such as $m=500$, when implemented in a doubly-recursive Horner form. This turns out to be quite convenient for this application, where we typically use grades of $15$ or so but sometimes as high as $40$.

Blends can be integrated exactly, as follows, and this is useful~\cite{corless2020blends}:
\begin{align}
\int_{s=0}^1 H_{m,n}(s)\,ds = &
{\frac { \left( m+1 \right) !}{ \left( m+n+2 \right) !}\sum _{j=0}^{m}
{\frac { \left( n+m-j+1 \right) !}{ \left( j+1 \right)
 \left( m-j \right) !}}}\,p_{{j}}\nonumber\\
 &+{\frac { \left( n+1 \right) !}{ \left( m+n+2
 \right) !}\sum _{j=0}^{n}{\frac {  \left( n+m-j
+1 \right) !}{ \left( j+1 \right)  \left( n-j \right) !}}}\,\left( -1 \right) ^{j}q_{{j}}
\>.
\label{eq:sumasintegral}
\end{align}
The numbers showing up in this formula turn out to be smaller for the higher-order Taylor coefficients, as one would expect.  Note that the above formula gives (in exact arithmetic) the \emph{exact} integral of the blend over the whole interval.  If the blend is approximating a function $f(s)$, then integrating equation~\eqref{eq:errorblend} gives us
\begin{equation}\label{eq:integrationerror}
  \int_{s=0}^1 f(s)\,ds - F(1) = (-1)^{n+1}\frac{(m+1)!(n+1)!}{(m+n+3)!} \frac{f^{(m+n+2)}(c)}{(m+n+2)!}
\end{equation}
where, using the Mean Value Theorem for integrals and the fact that $s^{m+1}(1-s)^{n+1}$ is of one sign on the interval, we replace the evaluation of the derivative at one unknown point $\theta$ with another unknown point $c$ on the interval.

Indeed, as described in~\cite{corless2020blends}, one can construct a new blendstring $H(z)$ for the antiderivative $F(z)$ from a blendstring $h(z)$ for $f(z)$, so that $H'(z) = h(z)$ exactly (up to roundoff error), and well approximates the antiderivative of $f(z)$.  This is useful for the problem at hand.

The code is available at \url{https://github.com/rcorless/Puiseux-series-Mathieu-double-points} in the files \texttt{ActiveLoopc1p0.maple} for the simple eigenvalue case and \texttt{ActiveDoubles.maple} for the double eigenvalue case.

\subsubsection{Marching\label{sec:HermiteObreschkov}}
We chose an implicit marching method based on Taylor series generation\footnote{Taylor series methods for solving IVP for ODE have historically been considered impractical by many people, but in fact this is not so, especially if the series coefficients can be generated easily, as in this case.  The quality of the free interpolants that one gets turns out to be a significant benefit. 
Taylor series methods have other benefits as well: see~\cite{nedialkov2005solving} and its references.}, quite standard in outline, as follows.  Taylor series coefficients that have been generated at the current node, say $\eta_n$, are supposed to be``known".  Specifically, suppose to start with that we have generated a Taylor polynomial of grade $m$ for our desired solution at this point.

Suppose also that we have chosen a tentative next node, $\eta_{n+1} = \eta_n + h$.  If our variable $\eta$ were time, this would be a time step.  The stepsize $h$ is tentative at this point.  We now generate Taylor coefficients for two independent solutions, satisfying (for one solution)
\begin{equation}
    y(\eta_{n+1}) = 1 \ \mathrm{and}\ y'(\eta_{n+1}) = 0
\end{equation}
and (for the complementary solution)
\begin{equation}
    y(\eta_{n+1}) = 0 \ \mathrm{and}\ y'(\eta_{n+1}) = 1\>.
\end{equation}

Next, we \textsl{blend} the known coefficients at $\eta_n$ with these independent solutions in the following way.
Form a blend of the known coefficients at $\eta_n$ with the \textsl{zero} Taylor series at $\eta_{n+1}$.  Call the result $L(\eta)$.  Form a blend of the first series above at $\eta_{n+1}$ with the \textsl{zero} Taylor series at $\eta_n$.
Call the result $C(\eta)$. Form a blend of the second series above with the \textsl{zero} Taylor series at $\eta_n$ and call the result $S(\eta)$.  Our desired solution will then be a linear combination of these three: say $y = A\,C(\eta) + B\,S(\eta) + L(\eta)$.  This uses the linearity of the equation, and the linear dependence of blends on their constituent Taylor coefficients.

We then use \textsl{collocation} at the two points $\eta_n + h/4$ and $\eta_n + 3h/4$ (which are Chebyshev--Lobatto points, not that it matters much at this low order) to give us two equations in the two unknowns $A$ and $B$.  That is, we compute the residuals
\begin{align}
   r_L(\eta) :=& L'' + (s - 2q\cos2\eta)L \nonumber\\
   r_C(\eta) :=& C'' + (s - 2q\cos2\eta)C \nonumber\\
   r_S(\eta) :=& S'' + (s - 2q\cos2\eta)S
\end{align}
at those two points, and set the residual for $y$ to zero at those two points:
\begin{align}
    0 = & A\, r_C(\eta_n + h/4) + B\, r_S(\eta_n+h/4) + r_L(\eta_n + h/4) \nonumber\\
    0 = & A\, r_C(\eta_n + 3h/4) + B\, r_S(\eta_n+3h/4) + r_L(\eta_n + 3h/4)\>.
\end{align}
We solve this two-by-two linear system by the exact formula for the inverse (this is as good a method as any, for such a small system) to acquire the coefficients $A$ and $B$.  This system is nonsingular because the solutions are linearly independent at the right endpoint and are well-scaled and well-conditioned in practice, as we observed experimentally.

Collocation is a well-understood technique for boundary-value problems for ODE~\cite{ascher1979colsys,ascher1995numerical}, but it has historically been used successfully for stiff initial-value problems as well~\cite{Wouk1976}.

After having computed $A$ and $B$ and used them to form our tentative solution $y$, we then \textsl{sample} the residual of $y$, namely $y'' + (s-2q\cos2\eta)y$, at the midpoint $\eta_{n+1/2} = \eta_n + h/2$.  This is (asymptotically as $h\to 0$) the location of the maximum residual over the step.  If this is smaller than our tolerance, we accept the step and continue.  Note that if the step is accepted, the Taylor coefficients then become known at $\eta_{n+1}$, being simply the known linear combination of the first and second sets of computed series coefficients.  We also use the measured residual (by known step-size control techniques~\cite{Gustafsson1988}) to predict the next step size $h_{n+1}$ and thus $\eta_{n+2}$.

If the step is rejected instead because it does not satisfy the accuracy tolerance, we reduce the stepsize by an amount indicated by the size of the measured residual (taking the order $2m$ into account), and try again.

Various known heuristics and safety factors are included in order to be cautious about various contingencies (for instance, the measured residual might be accidentally small, which throws the predicted stepsize off; similarly, the stepsize predictions are determined by assuming that the derivatives involved in the error coefficients ``don't change much'' from step to step, but this is sometimes violated in practice).  Error messages can be generated if too many stepsize reductions are encountered, or if the solver can't find a good starting stepsize\footnote{We start with a pure Taylor series to estimate the initial step size $h$.  This has some potential to go wrong, and sometimes does, because it does not benefit from implicitness, but we have found it satisfactory.}, or if the maximum number of steps is reached, as is usual with IVP solvers.

\subsubsection{Rationale\label{sec:rationale}}
The \textsl{reasons} we do this, instead of using a more standard method that has already been implemented and tested, include the following.
\begin{enumerate}
\item We work from the beginning over the complex plane (most standard implementations put integration over the real line first).
\item We can handle the double-eigenvalue case in a straightforward way.  To be fair, other methods can also handle this case in a straightforward way, as well, but at least we are not at a disadvantage.
\item The functions are entire, and therefore Taylor series are defined everywhere for them. Since blendstrings are very smooth (with grade $m$ Taylor coefficients at each knot, they are $m$ times continuously differentiable) they may be expected to be accurate and convenient.
\item The problem is linear, so the implicitness of the method is simple to deal with (and there are no convergence issues in solving nonlinear equations at each step).
\item Putting $\eta = \eta_n + sh$, the residual has the error expression\footnote{To show this, notice that the residual is $O((z-\alpha_k)^{m-1})$ at the left endpoint (not $O((z-\alpha_k)^{m+1})$ because we have differentiated twice), is $O((z-\alpha_{k+1})^{m-1})$ at the right endpoint, and vanishes at the Chebyshev--Lobatto points in between.}
\begin{equation}
    r(\eta) = K h^{2m} s^{m-1}(s-\frac14)(s-\frac34)(1-s)^{m-1} + O( h^{2m+1} )
\end{equation}
for some ``constant" $K$ depending on high-order derivatives of the solution, evaluated at some point in the interval. In comparison to an explicit Taylor series method, this gains a factor of $2^{2m+2}$ in accuracy because the maximum value of the polynomial in $s$ is $2^{-2m-2}$.  Since we typically take $m=15$ or higher, this accuracy gain is noticeable.
\item The effect of the residual on the solution can be analyzed by using the Green's function for the Mathieu equation, which can easily be computed by the same methods:
\begin{equation}
    G(\eta,\tau) = w_I(\eta)w_{II}(\tau) - w_I(\tau)w_{II}(\eta)\>.
\end{equation}
Here we use the notation for the basic solutions as described in the DLMF~\url{https://dlmf.nist.gov/28.2.ii}.
The change in solution produced by a residual $r(\eta)$ is 
\begin{equation}
    \int_{\tau=0}^\eta G(\eta,\tau) r(\tau)\,d\tau\>.
\end{equation}
\item We can re-use standard stepsize heuristics, which are well-known to produce ``good" meshes which reflect dynamic changes in the solution.
\item The Mathieu equation is not ``stiff" with the stepsizes and tolerances we are using~\cite{soderlind2015stiffness}, but is rather \textsl{oscillatory}, and as such benefits somewhat from the implicitness of this method.  There is still a stability restriction, but it is not very important compared to the stepsize restriction needed for accuracy, and this implicit method does perform better than a pure explicit Taylor series method.
\item Using a residual (defect) control is useful even for \textsl{unstable} differential equations.  The modified Mathieu equation can be very unstable, exhibiting doubly exponential growth.
\item These Taylor coefficients are very easy to generate, and the code is quite simple.  The fact that the order of accuracy can be chosen more or less arbitrarily is an advantage for very high-precision computation: the cost for accurate solution is \textsl{polynomial} in the number of bits of accuracy~\cite{ilie2008adaptivity}.
\item We do want high-precision computation, because we want to be able to state unequivocally that numerical artifacts are not present, and to verify that any given solution is as accurate as the code claims.  This is not a given, without an external check, because of the heuristics and safety factors needed in practice for the solver.
\item Taking derivatives and integrals of blends is very simple, and both of these are needed for subsequent computations with the solution.  We are not just interested in the solution, but also in integrals and derivatives of the solution.
\item It \textsl{might} be true that this method is useful for the numerical solution of some other, similar, equations.  In particular this might be of interest for D-finite (holonomic) systems.  This application provides a useful test case.
\end{enumerate}

\subsection{Testing the numerical solution}
Because each computed Mathieu function and modified Mathieu function is a smooth piecewise polynomial, it can be differentiated and substituted back into the differential equation. What is left over is sometimes called the ``defect" but the more usual name in numerical analysis is the \textsl{residual}. The solutions always had a residual comparable to the tolerance with which the solver was called; typically about $10^{-11}$ if we were working in double precision, and about $10^{-28}$ if we were working in $30$ decimal digits.  This is, of course, not enough to say that the forward error is small: one needs also to compute the Green's function, or otherwise verify that the condition number\footnote{By this we do not mean the condition number of a matrix (there are no matrices here) but rather the condition number of the Mathieu differential equation, which since the equation is linear, is equivalent to the maximum value of the Green's function.} is small.

It turns out that for $\fa$ near $1$, i.e. when the ellipse is nearly circular and the eccentricity $\eccentricity$ is small, then $\xi_0$ gets modestly large---and because the modified Mathieu functions grow doubly exponentially, the Green's function does indeed amplify errors in this case. Indeed, the condition number of $\Ce_0(\xi,q)$ for evaluation, namely $C = \xi \Ce_0'(\xi,q)/\Ce_0(\xi,q)$ always grows exponentially with $\xi$.  For $q = 0.3823i$, a typical value of the parameter, the value of the condition number $C(4.0)$ is approximately $120$. This is tolerable.  For all values of the parameters that we used, except for the stress test when $\fa = 0.9999$ (more about this, below), the condition number was similarly modest.

Another way to see this is to vary the parameters (such as $\fa$ or $\omega$) and verify that the solution does not change much.  A third way is to do the computations again in higher precision.  We found in the end that our computation of the Mathieu functions and modified Mathieu functions was very reliable.

\begin{figure}
    \centering
    \begin{subfigure}[b]{0.4\textwidth}
         \centering
         \includegraphics[width=0.9\textwidth]{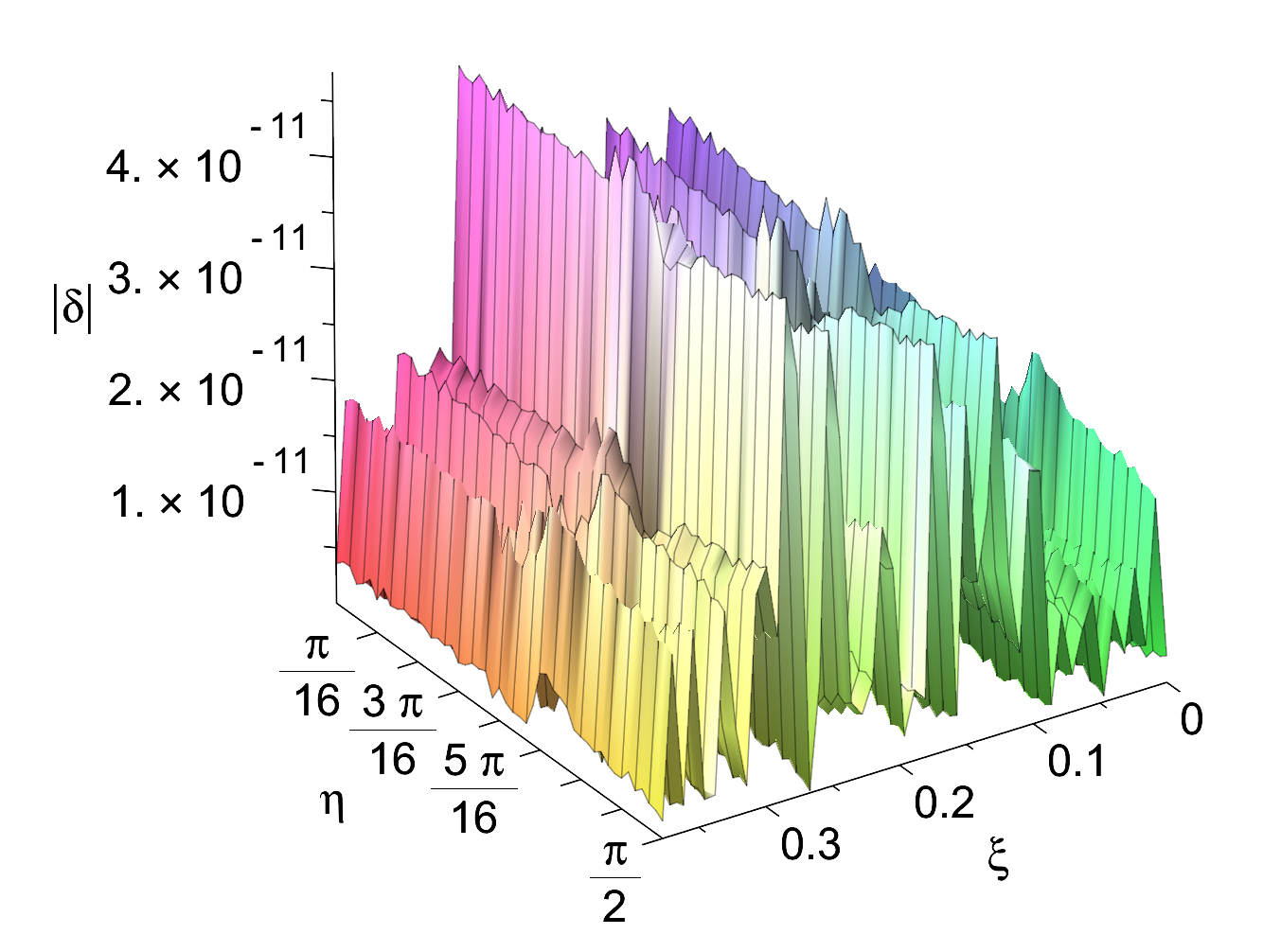}
         \caption{Absolute value of the residual}
         \label{fig:absresidual}
    \end{subfigure}
    \hfill
    \begin{subfigure}[b]{0.4\textwidth}
    \centering
    \includegraphics[width=0.8\textwidth]{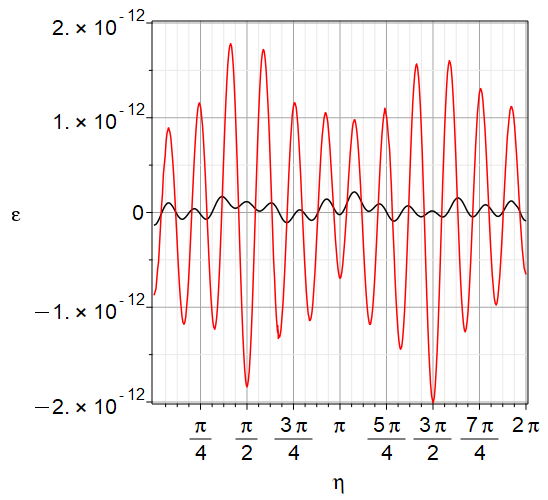}
    \caption{Truncation error at the boundary.}
    \label{fig:boundaryerror}
    \end{subfigure}
    \caption{Guarding against blunders and errors: On the left, the computed residual $\delta = \partial^2v/\partial \xi^2 + 
    \partial^2 v/\partial \eta^2 - i\rho\omega d^2 v/(2\mu)$ with $N=6$ terms, $f=0.6$, $c=1.0$cm, active scenario, $\omega = 3.83722019332829$ (a double eigenvalue case), in double precision.  This shows that the differential equation is satisfied to better than $10^{-10}$.  On the right, we plot our computed $\epsilon = k_0(v(\xi_0,\eta)-1)/(i\rho \omega)$, which ideally should be zero at the boundary, for the same parameter values, real part in black and imaginary part in red, on $0 \le \eta \le 2\pi$.  We see the effect of truncating our series expansion at $N=6$. Because this is a double eigenvalue case and the Green's function was used, the piecewise polynomial approximation for the solution is not \textsl{quite} periodic: there is a tiny jump between the values at $\eta=0$ and at $\eta=2\pi$. Indeed the error is not quite periodic with period $\pi$, which it would be ideally. }
\end{figure}

It is also possible to verify that the underlying PDE is satisfied: one computes (for instance) $\partial^2 w/\partial \xi^2$,   $\partial^2 w/\partial\eta^2$, and $w$ at one or several or a great many points, and substitues these values back into the partial differential equation~\eqref{eq:governingEquationChanged}. When we do this we see that what is left over is about the size of the integration tolerance (typically quite near to the unit roundoff level in our runs), over the whole ellipse.    Figure~\ref{fig:absresidual} shows the result of doing this in one case, for $v(\xi,\eta)$ using equation~\eqref{eq:equationInV}.  This kind of \textsl{a posteriori} solution validation is a powerful check against numerical errors.  What we have proved by this a posteriori computation is that we have computed the exact solution of a perturbed PDE, where the perturbation is smaller than $10^{-10}$.  Compared to modelling errors (for instance that the true deformed shape is not exactly elliptical, or the even greater modelling error of neglecting the third dimension) this shows definitively that the numerical method has performed satisfactorily. This is a useful guard against blunders, as well: We were reminded to use the chain rule, and also found a typographical error in one equation, when we did this.

Finally one needs to check the boundary conditions.  In figure~\ref{fig:boundaryerror} we see one such check.  The oscillatory nature of the error indicates that it is truncation error we are seeing---the effect of taking only $N=6$ terms in the expansion.  With a high enough $N$, one sees only rounding errors at this stage.

\subsubsection{Difficult cases for the code}
If $q$ is \textsl{small}, then the continued fraction approach of Blanch becomes somewhat fragile.  Blanch had performed a good numerical analysis of the method, and using her methods it can be made to work well in this situation by various adjustments. In contrast, however, the performance of the matrix method \textsl{improves} as $q \to 0$, so it is simpler just to drop the use of the continued fraction approach when $q$ is ``too small".  We chose, somewhat arbitrarily, to use just the matrix method if $|q| < 0.2$.

The solver is meant for use by people willing to adjust parameters experimentally and not, in Blanch's words, simply to be run ``in a robot-like fashion." The code has, in particular, an aggressive initial step-size heuristic based on explicit Taylor series.  This got into trouble for some of the runs in the passive scenario with circumference $2.0$cm and fractions $\fa$ closer to $1$.  We could have adjusted the parameters (tolerance and grade $m$ of Taylor approximation) but it was simpler to use higher precision for those runs, which we did in $30$ decimal digits.  The time penalty was slight, even though we have not optimized the code for speed.  The first step for that, of course, would be to use a production language instead of a prototyping language such as Maple (which saves our time and not the computer's).  Even so, the solver is gratifyingly rapid, even at very high precision.

The only real difficulty that occurs with expansion in Mathieu functions is when the fraction $\fa$ is nearly~$1$.  That is, the nearly-circular case is the difficult one for expansion in terms of Mathieu functions.  This is because the coordinate transformation used, namely $x=d\cosh\xi\cos\eta$ and $y=d\sinh\xi\sin\eta$, becomes singular as the focal distance $d \to 0$, which it must as the ellipse becomes a circle.  Another way to think about this is to ``zoom out" on confocal ellipses; the larger the diameter, the more nearly circular the confocal ellipses are.

This singular behaviour shows up in several ways, numerically.  For instance, taking $\fa=0.999$ in the active scenario, and choosing $\omega$ so $q$ is the Mulholland--Goldstein point, seems that it should not cause problems.  But it does, because $\xi_0 = \mathrm{invsech}(\eccentricity)$ is then about $4.6$. Although that does not seem like much, the modified Mathieu functions grow \textsl{doubly exponentially}\footnote{One is used to exponential growth, but doubly exponential growth is remarkably difficult to deal with.  To see the asymptotics for $\Ce(\xi,q)$ look at the DLMF.  See~\url{https://dlmf.nist.gov/28.25} in particular.}; in this case, $\Ce_0(\xi_0,q^*)$ has magnitude $6.0972\cdot 10^{45}$.  The modified Mathieu equation becomes very difficult to integrate accurately for large $\xi$ because of this doubly exponential growth.

As a stress-test for the code, we solved the problem at very high precision with $\fa=0.9999$ in both the active and passive scenarios (which wind up being very similar, of course).  Using $200$ decimal digits of precision, and Taylor series of degree $100$ (so the numerical method was of order $200$) we were able to solve the problems accurately in only a few seconds.  It is ironic that the ``difficult case" resembles so strongly the simple case of a tube of circular cross section, which has a direct and natural solution in terms of Bessel functions.

\subsubsection{Comparison with a standard code}
When we compare this code with that of~\cite{Shampine2000}, we see that that standard code performs very well, in fact. Even just with the default solver (a version of RKF45) all scenarios are rapidly solved.  The doubly-exponential growth of the modified Mathieu equation is simply taken in stride by the code.  We remark that that code, while more than 20 years old, has undergone steady development since then at the hands of Allan Wittkopf of Maplesoft, who has (without publishing papers on the subject) incorporated many speed and reliability enhancements.

However, access to the internal interpolants used by the code is quite awkward, and it is not easily possible to differentiate the interpolant to compute a residual to \textsl{validate} the solution it produces. With the present code, this is simple (indeed automatic).  Secondly, if very high precision is wanted, the higher order of the present code lowers the cost.  Indeed, using this method, the cost of solution is polynomial in the number of bits of accuracy requested~\cite{ilie2008adaptivity}, while for fixed-order Runge--Kutta methods the cost is exponential in the number of bits of accuracy requested.  At modest accuracy, or even at double precision accuracy, this is not a problem, of course.

The third advantage of the present method is the decent numerical properties of the underlying interpolant.  In comparison, the monomial basis used by the internal Maple code can suffer more from rounding errors (at high precision), although we have no doubt that the developers have taken steps to minimize the difficulty.

Something that might have been a fourth advantage, the ease of combining and integrating blends to compute, for instance, the Green's function, is not much of an advantage after all: Maple's \lstinline{dsolve/numeric} interface has several flexible features that let one combine solutions, and integrating the solution of a differential equation is merely a matter of integrating the differential equation for the integral in question.

Still, this present code offers some potential advantages for other applications, and using this problem as a test case for it has proved to be interesting.

\section{The double-eigenvalue case\label{sec:doubleeig}}

In the case of a double eigenvalue, the previous formulae need to be amended.  For the Mathieu equation, double eigenvalues are isolated, and there are no higher-order eigenvalues, so the treatment is relatively straightforward.  The theory has been known for a long time~\cite{meixner}, but in practice it seems to have been ignored.  We therefore give a detailed treatment below.

We will use a Puiseux series expansion near the double eigenvalue to deduce the analytical form needed for expansion exactly at the double point. We emphasize that the computations in this section are exact computation of series coefficients, and analytical cancellation of large terms will give us the result that we want.

Now suppose that the coalescing eigenfunctions are $\ce_0(\eta,q)$ and $\ce_2(\eta,q)$.  For our computations this is the one that mattered the most, when $q \approx 1.4668\, i$ is the Mulholland--Goldstein point; but other purely imaginary eigenvalues also occur for larger frequencies, or larger circumference blood vessels; so we give details of the process.

If $q^* = 1.4687686137851\ldots\,i$ is the Mulholland--Goldstein point, then we may expand the eigenvalues $a_0$ and $a_2$ in Puiseux series to get $a_0 = a^* - \alpha_1 \sqrt{q-q^*} + O(q-q^*)$ and $a_2 = a^* + \alpha_1 \sqrt{q-q^*} + O(q-q^*)$ in a region close to that point.  Here $a^* = 2.08869890274969540\ldots$ is real, and known to many decimal places~\cite{Brimacombe2021}. Similarly, 
\begin{equation}
    \alpha_1 = 1.659487804320\ldots + 1.659487804320\ldots\,i
\end{equation}
is also known to many decimal places, although as we will see it does not appear in the final formulae for the spectral expansion coefficients at the double eigenvalue.

If we write a Mathieu series expansion for some function, say $v(\xi,\eta)$, at a point \textsl{near} to the double point, we find that the coefficients of the terms $\Ce_0(\xi,q)\ce_0(\eta,q)$ and $\Ce_2(\xi,q)\ce_2(\eta,q)$ are large and of opposite sign; indeed they have leading behaviour that is $O((q-q^*)^{-1/2})$.  Also, all of $\Ce_0$, $\ce_0$, $\Ce_2$, and $\ce_2$ can be written as functions of the fundamental solution $w_I(z,a,q)$ as follows:
\begin{align}
\ce_0(\eta,q) =& w_{I}(\eta, a_0, q ) \nonumber \\
\Ce_0(\xi,q)  =& w_{I}(i\xi, a_0, q ) \nonumber \\
\ce_2(\eta,q) =& w_{I}(\eta, a_2, q ) \nonumber \\
\Ce_2(\xi,q) =& w_{I}( i\xi, a_2, q ) \>.
\end{align}
We will also need the following two new functions:
\begin{align}
    u(\eta) :=& D_2(w_I)(\eta,a^*,q^*) \label{eq:generalizedeigenfunction}\\
    U(\xi)  :=& D_2(w_I)(i\xi, a^*, q^*)\>. \label{eq:complementarygeneralizedeigenfunction}
\end{align}
Here $D_2(f)(x,y,z)$ means the partial derivative with respect to the second variable, and then evaluated at the point $(x,y,z)$. We will show in a following subsection how these can be computed.

Now suppose that the solution at the point $q$ has the expansion
\begin{equation}
    v(\xi,\eta) = b_0 \Ce_0(\xi,q)\ce_0(\eta,q) + b_2 \Ce_2(\xi,q)\ce_2(\eta,q) + \cdots 
\end{equation}
where the terms not included have eigenvalues that will not coalesce, and therefore the previous treatment using orthogonality will suffice to identify their coefficients $b_{2m}$ for $m>1$. 
Putting for brevity $x = \sqrt{q-q^*}$ and expanding everything in series in $x$ and neglecting terms of size $O(x^2)$ or smaller, we have the following:
\begin{align}
    a_0 =& a^* - \alpha_1 x + O(x^2) \nonumber \\
    a_2 =& a^* + \alpha_1 x + O(x^2) \nonumber \\
    b_0 =& \frac{A_0}{x} + B_0 + O(x) \nonumber \\
    b_2 =& \frac{A_2}{x} + B_2 + O(x) \nonumber \\
    \Ce_0(\xi,q) =& \Ce_0(\xi,q^*) -\alpha_1 U(\xi) x + O(x^2) \nonumber\\
    \ce_0(\eta,q) =& \ce_0(\eta,q^*) -\alpha_1 u(\eta) x + O(x^2) \nonumber\\
    \Ce_2(\xi,q) =& \Ce_0(\xi,q^*) + \alpha_1 U(\xi) x + O(x^2) \nonumber\\
    \ce_2(\eta,q) =& \ce_0(\eta,q^*) + \alpha_1 u(\eta) x + O(x^2) \>.
\end{align}

Now in our case, the coefficients of $v(\xi,\eta)$ are determined by integration against the constant function $1$ at the wall $\xi=\xi_0$, so that for $q$ near to $q^*$ we have
\begin{equation}
    \int_{\eta=0}^{2\pi} \ce_0(\eta,q^*)\,d\eta - \alpha \int_{\eta=0}^{2\pi} u(\eta)\,d\eta x + O(x^2)
\end{equation}
on the left-hand side, and, expanding everything out and using the fact that 
\begin{equation}
\int_{\eta=0}^{2\pi} \ce_0^2(\eta,q^*)\,d\eta = 0\>,
\end{equation}
we find
\begin{equation}
    -2\alpha_1 A_0 \Ce_0(\xi_0,q^*) \int_{\eta=0}^{2\pi} \ce_0(\eta,q^*)u(\eta)\,d\eta + L_1 x + O(x^2)
\end{equation}
on the right-hand side, with
\begin{equation}
    L_1 = \alpha_1^2 A_0 \Ce_0(\xi_0,q^*)\int_{\eta=0}^{2\pi} u^2(\eta)\,d\eta 
    -2\alpha_1\left(B_0\Ce_0(\xi_0,q^*) -\alpha_1 A_0U(\xi_0) \right)\int_{\eta=0}^{2\pi}\ce_0(\eta,q^*)u(\eta)\,d\eta\>.
\end{equation}
Since the squared integral of the generalized eigenfunction $u(\eta) = D_2(w_I)(\eta,a^*,q^*)$ is not zero, and since the integral of the product of $u(\eta)$ with $\ce_0(\eta,q^*)$ is not zero, and since $\Ce_0(\xi_0,q^*)$ is not zero, we may equate the constant terms and the terms linear in $x$ and solve for $A_0$ and for $B_0$.  We get
\begin{equation}
    A_0 = -\frac{ \int_{\eta=0}^{2\pi} \ce_0(\eta,q^*)\,d\eta}{2\alpha_1\Ce_0(\xi_0,q^*) \int_{\eta=0}^{2\pi} \ce_0(\eta,q^*)u(\eta)\,d\eta}
\end{equation}
and
\begin{equation}
    B_0 = \frac{\int_{\eta=0}^{2\pi} u(\eta)\,d\eta + \alpha_1A_0\left(2U(\xi_0)\int_{\eta=0}^{2\pi} \ce_0(\eta,q^*)u(\eta)\,d\eta + \Ce_0(\xi_0,q^*)\int_{\eta=0}^{2\pi} u^2(\eta)\,d\eta\right)}{2\Ce_0(\xi_0,q^*)\int_{\eta=0}^{2\pi} \ce_0(\eta,q^*)u(\eta)\,d\eta}
\end{equation}
Similarly, we get
\begin{equation}
    A_2 = -A_0 =  \frac{ \int_{\eta=0}^{2\pi} \ce_0(\eta,q^*)\,d\eta}{2\alpha_1\Ce_0(\xi_0,q^*) \int_{\eta=0}^{2\pi} \ce_0(\eta,q^*)u(\eta)\,d\eta}
\end{equation}
and
\begin{equation}
    B_2 = \frac{\int_{\eta=0}^{2\pi} u(\eta)\,d\eta - \alpha_1A_2\left(2U(\xi_0)\int_{\eta=0}^{2\pi} \ce_0(\eta,q^*)u(\eta)\,d\eta + \Ce_0(\xi_0,q^*)\int_{\eta=0}^{2\pi} u^2(\eta)\,d\eta\right)}{2\Ce_0(\xi_0,q^*)\int_{\eta=0}^{2\pi} \ce_0(\eta,q^*)u(\eta)\,d\eta}
\end{equation}
which resembles the formula for $B_0$.

Putting these formulae into the expansion for $v(\xi,\eta)$ we get $v(\xi,\eta) = (B_0+B_2)\Ce_0(\xi,q^*)\ce_0(\eta,q^*) - 2\alpha_1A_0\left( U(\xi)\ce_0(\eta,q^*) + \Ce_0(\xi,q^*)u(\eta)\right) + O(x)$ and thus as $q \to q^*$ the expansion of $v(\xi,\eta)$ becomes
\begin{equation}
    v(\xi,\eta) = b_0\Ce_0(\xi,q^*)\ce_0(\eta,q^*) + \hat{b}_2\left(U(\xi)\ce_0(\eta,q^*) + \Ce_0(\xi,q^*)u(\eta) \right) + \cdots
\end{equation}
where
\begin{align}
b_0 =& \frac{\int_{\eta=0}^{2\pi} u(\eta)\,d\eta}{\Ce_0(\xi_0,q^*)\int_{\eta=0}^{2\pi} \ce_0(\eta,q^*)u(\eta)\,d\eta} \nonumber \\
& + \frac{\left(\int_{\eta=0}^{2\pi}\ce_0(\eta,q^*)\,d\eta\right)\left(\Ce_0(\xi_0,q^*)\int_{\eta=0}^{2\pi} u^2(\eta)\,d\eta + U(\xi_0)\int_{\eta=0}^{2\pi} \ce_0(\eta,q^*)u(\eta)\,d\eta\right)}{\left(\Ce_0(\xi_0,q^*)\int_{\eta=0}^{2\pi} \ce_0(\eta,q^*)u(\eta)\,d\eta\right)^2}
\end{align}
\begin{equation}
\hat{b}_2 =  \frac{\int_{\eta=0}^{2\pi} \ce_0(\eta,q^*)\,d\eta}{\Ce_0(\xi_0,q^*)\int_{\eta=0}^{2\pi} \ce_0(\eta,q^*)u(\eta)\,d\eta} \>,  
\end{equation}
and the omitted terms can all be calculated by orthonormality as before. 

\begin{remark}
    This result can be derived a different way, by differentiating the original formula with respect to $a$.  With that method, the appearance of $U(\xi)\ce_0(\eta) + \Ce_0(\xi)u(\eta)$, being the derivative of $\Ce_0(\xi)\ce_0(\eta)$, seems natural. Then one can use the orthogonality of $\ce_0(\eta)$ with all $\ce_{2m}(\eta)$ (including itself) to compute $\hat{b}_2$, and then integrate against $u(\eta)$ and solve the resulting equation for $b_0$ using the known $\hat{b}_2$.  This leads to the same result, but we feel that the detailed derivation above is more convincing, and explains what happens to the expansion coefficients as $q \to q^*$.
\end{remark}
All that remains is the computation of $u(\eta)$ and $U(\xi)$.  To do this, we compute the Fr\'{e}chet derivatives of the Mathieu equation and the modified Mathieu equation:
\begin{align}
    \frac{d^2 u}{d\eta^2} + (a^* - 2q^*\cos2\eta) u + y &= 0 \label{eq:FrechetMathieu} \\
    \frac{d^2 U}{d\xi^2} - (a^* - 2q^*\cosh 2\xi) U - y &= 0 \>.\label{eq:FrechetModifiedMathieu}
\end{align}
In the first equation, replace $y$ by $\ce_0(\eta,q^*)$ and solve (we use the Green's function for the Mathieu equation to do so, because algebraic operations and integration are accurate and efficient with blendstrings) and similarly in the second equation replace $y$ by $\Ce_0(\xi,q^*)$ and solve.  As for initial or boundary conditions, we need to take $u(0) = u(2\pi) = 0$ to ensure periodicity, and we need to take $U(0) = U'(0) = 0$ to ensure symmetry at the line $\xi=0$. 

This analysis is implicit in the treatment in~\cite{meixner}, but does not seem to be widely pursued, and so we have written it down in some detail here.

Finally, we must amend the formulas for flow rate and oscillatory wall shear stress.  Equation~\eqref{eq:oscillatoryVolumetricFlow} becomes

\begin{align} \label{eq:modifiedoscillatoryVolumetricFlow}
q_{\phi,e}(t) = \frac{8q_{0,e}}{i \lambda_{e}} \Big( 1 - \frac{1 }{i \pi \lambda_{e}} \tanh 2 \xi_{0} \Big( & 
b_0  \Ce_0'(\xi_0,q^*) \int_{\eta=0}^{2\pi} \ce_0(\eta,q^*) 
\nonumber \\ & + \hat{b}_2\left(U'(\xi_0)\int_{\eta=0}^{2\pi}\ce_0(\eta,q^*)\,d\eta + \Ce_0'(\xi_0,q^*)\int_{\eta=0}^{2\pi} u(\eta)\,d\eta\right)  \nonumber\\
& +\sum_{m \ge 2} b_{2m}^2 I_{2m} \Ce_{2m}'(\xi_0,q^*)\Ce_{2m}(\xi_0,q^*) \Big)\Big)e^{i \omega t} \>.
\end{align}
The integrals appearing in the formula above have already been calculated, in order to find the $b_{2m}$, but the relationships used to simplify to get $b_{2m}^2 I_{2m}$ as in the rest of the sum no longer obtain because $I_0 = 0$.

Equation~\eqref{eq:oscillatoryShearStress} becomes
\begin{equation} \label{eq:modifiedoscillatoryShearStress}
\tau_{\phi,e} \left( \eta, t \right) =
\frac{2  \hat{\tau}_{0,e} }{i\delta_0 \lambda_{e}} \left(
\hat{b}_2 \left(U'(\xi_0)\ce_0(\eta,q^*) + \Ce_0'(\xi_0,q^*)u(\eta)\right) + 
\sum_{m\ne 1} {b}_{2m} \Ce_{2m}'(\xi_0,q^*)\ce_{2m}(\eta,q^*)\right)e^{i \omega t} \>.
\end{equation}

\subsection{The value of being able to solve the double eigenvalue case}
Because the solution to the original model equations is continuous (indeed analytic) in the parameters involved in $q = -i\Lambda_e/4$, the underlying solution \textsl{changes continuously} as $q$ passes through a value where a double eigenvalue of the Mathieu equation occurs. Therefore, it is only a discontinuity in the \textsl{representation} of the solution, not the solution itself.  This means that sampling $q$ ``near enough'' to the double point would give solutions that are ``near enough'' to the solution at that point.

The only difficulty, and this is rather mild, is that the expansion coefficients in Mathieu functions become large and of opposite sign, which \textsl{might} incur some visible rounding error owing to cancellation.  Because  
the size of the coefficients is only $O(q-q^*)^{-1/2}$ this is not typically very severe.

Nonetheless we feel that it is worthwhile to be able to give the precise solution exactly at a double point for comparison to simple solutions nearby, to be assured that the solutions shown are representative of the model.
\section{Results and discussion}
In the results to follow we consider a change in the cross section of a tube from circular to elliptic, under both passive and active scenarios, and examine the effects of this on the properties of oscillatory flow in a tube of elliptic cross section under the same oscillatory pressure gradient as that in a tube of circular cross section.  The effects of physiological interest are those on flow rate and on the distribution of shear stress around the circumference of the tube. The main focus of our study is therefore on these two properties as well on the form of the passive scenario of velocity profiles in a tube of elliptic cross section.

As noted, the properties of oscillatory flow in a tube of elliptic cross section depend on the nondimensional parameter $\Lambda_{e}$ (Eq.~\ref{eq:LambdaDef}) which  involves the frequency of oscillation, $\omega$, as well as the focal distance, $d$. Thus the effects of tube dimension on oscillatory flow in the tube of elliptic cross section are different at different frequencies and, similarly, the effects of frequency on oscillatory flow in the tube of elliptic cross section are different at different tube dimensions. As a consequence, \textsl{the effects of tube dimension and of frequency cannot be scaled out} and, in the results to follow we examine three specific values of the circumference and thus radii, deformed by forcing them to different fractions $\fa$ of their original radii, and several specific values of frequency, as shown in the figures. 

All the results to follow are based on the real part of the oscillatory pressure gradient (equation~\eqref{eq:oscillatoryPressureGradient}). 
We note that the absolute value of the various complex quantities must appear at some point in the flow, possibly with a different phase lag for different locations in the vessel. In our animations (not given here) the differences in phase lags were never very significant.

The fluid density and viscosity in all the results were taken as 1.0 g/cm$^{3}$ and 0.04 g/(cm$\cdot$s), respectively. We note that we only examine tubes equivalent to those of radius of $0.5$cm, $1.0$cm, and $2.0$cm in both active and passive scenarios for maximum flow rate and maximum wall shear stress.

The primary factor in the transition of pulsatile blood flow in a vessel of circular cross section to one in a vessel of an elliptic cross section is the loss of radial symmetry of the circular cross section.
While from a geometrical perspective this loss of symmetry appears to occur fairly smoothly, from both a mathematical and a hemodynamical perspective it represents a significant change. Geometrically, the change from a circular to an elliptic cross section, however small, 
introduces ``poles'' in the cross section, places where the curvature is maximum---at the ends of the major axis---and where the curvature is minimum, at the ends of the minor axis.

Further, the most convenient coordinate system, namely confocal elliptical coordinates, has singular behaviour in the limit as the focus distance $d \to 0$. This in turn induces a change in the governing equations of pulsatile flow from Bessel equations to Mathieu equations. Hemodynamically, the change causes a redistribution of shear stress on the vessel boundary, from a uniform distribution in the case of circular cross section to a polarized distribution in the case of elliptic cross section, with maximum shear occurring at the two ends of the minor axis of the ellipse and minimum shear at the two ends of the major axis.

From the perspective of blood flow regulation, which our study was aimed at, the transition from flow in a vessel of circular cross section to one in an elliptic cross section represents a departure from well known physiological rules of blood flow regulation to a somewhat uncharted territory. A simple change in the diameter of a vessel is well known as the physiological (neurovascular) mechanism used to change the cross sectional area of a blood vessel in order to affect a required change in flow rate.

Our study was aimed at the question of how this well established rule of blood flow regulation might be altered in the case of an elliptic cross section. 
Our results indicate that if the regulatory system does not respond to the change from circular to elliptic cross section, which we have dubbed as a ``passive scenario'', the change from circular to elliptic cross section will occur with no change in the  length of circumference of the changing cross section. As a consequence, the cross sectional \textsl{area} available to the flow will then be reduced under this scenario.

If, on the other hand, the regulatory system intervenes in an attempt to maintain the cross sectional area available to the flow, as it does in the case of a circular cross section, the transition from circular to elliptic cross section will occur while keeping the cross sectional area constant and hence, necessarily, by increasing the length of circumference of the changing cross section. We have dubbed this as an ``active scenario''.

This makes a difference even in \textsl{steady} flow in tubes of elliptic cross section, as already seen in Figures~\ref{fig:steadyvelocityactiveredpassiveblack}--\ref{fig:steadyflowrate}.  The flow quantities tend to be higher under the active scenario.  This persists for pulsatile flow, as we will see; but we measure the pulsatile flow quantities relative to their steady counterparts, and so the increase may be hidden.  This must be kept in mind.

\begin{figure}[t]
     \centering
     \begin{subfigure}[b]{0.4\textwidth}
         \centering
         \includegraphics[width=\textwidth]{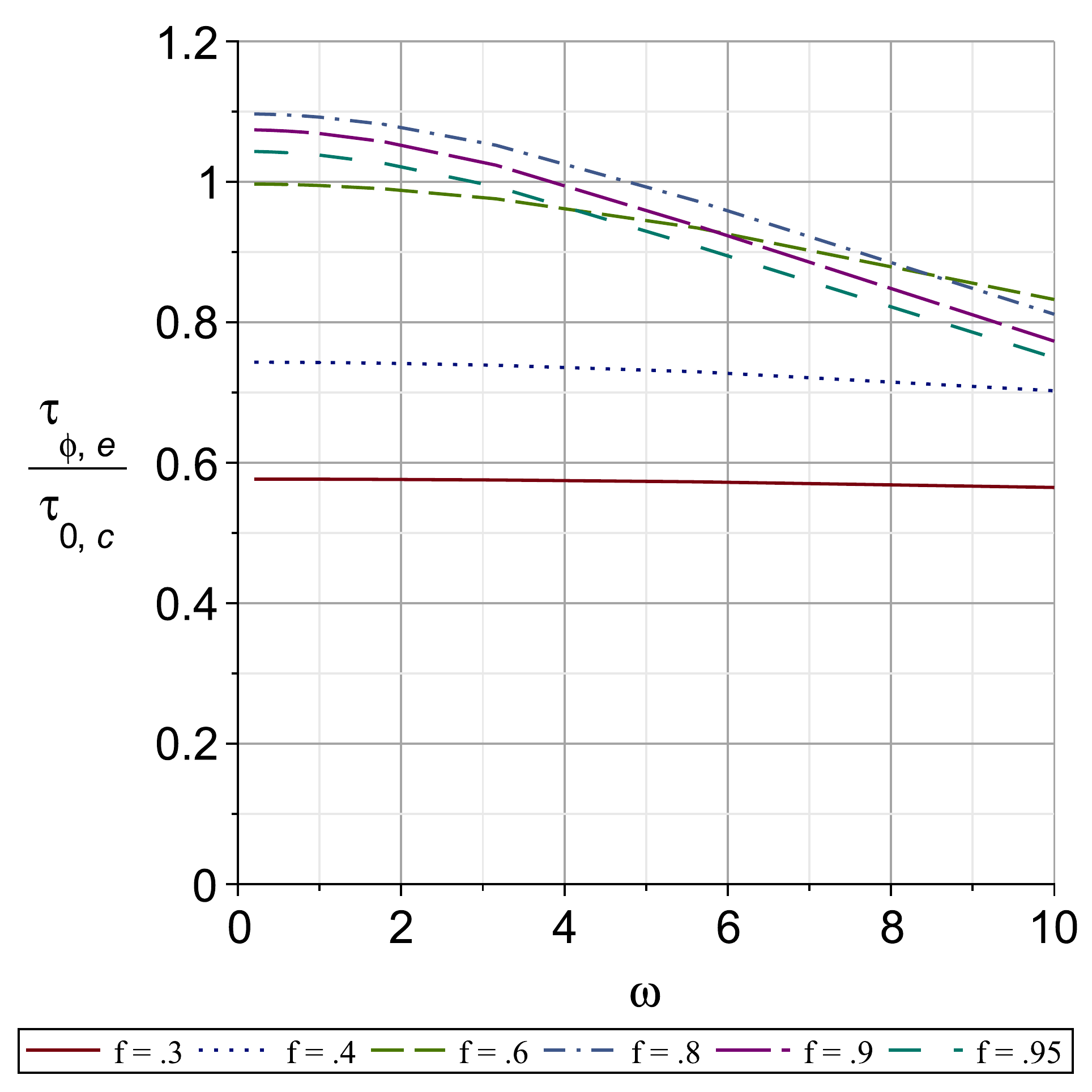}
         \caption{Passive, $c = 1.0$cm}
         \label{fig:PassivemaxwallshearStressc1p0}
     \end{subfigure}
     \hfill
     \begin{subfigure}[b]{0.4\textwidth}
         \centering
         \includegraphics[width=\textwidth]{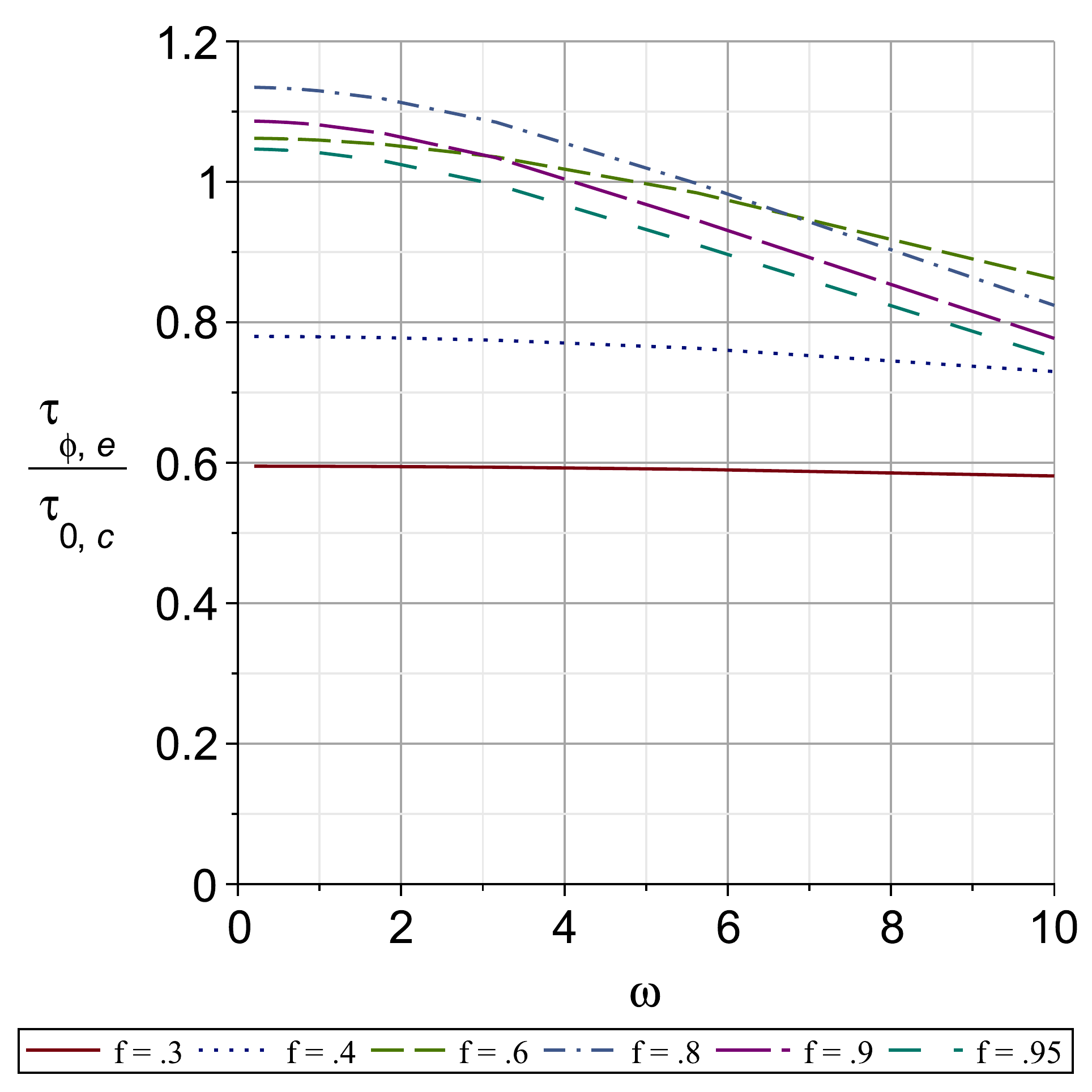}
         \caption{Active}
         \label{fig:ActivemaxwallshearStressc1p0}
     \end{subfigure}
        \caption{Maximum oscillatory wall shear stress from equation~\eqref{eq:oscillatoryShearStress} compared in the passive scenario (left) to the active (right). The initial circumference was $c=1.0$cm. The same range of fractions $\fa$ are graphed for each scenario, namely $0.3$, $0.4$, $0.6$, $0.8$, $0.9$, and $0.95$ over the same range of frequencies $\omega$.In this figure we see that the different scenarios produce very similar maximum wall shear stresses, and moreover that the different scenarios show a similar dependence on the imposed frequency $\omega$, with a general decrease in shear stress for higher frequency. This dependence is weaker with the more compressed cross sections (smaller fractions $\fa$).  We used the simple symbol $f$ instead of $\fa$ in the legend to improve readability.}
        \label{fig:ActiveVsPassivemaxwallshearStressc1p0}
\end{figure}

While what has been said so far applies to steady flow, the effects of ellipticity in oscillatory flow are further complicated by the acceleration and deceleration of the fluid within the oscillatory cycle. The effects of acceleration and deceleration depend on the volume of fluid being accelerated and decelerated, which in turn depends on the cross sectional area of the tube in which the flow is taking place. We recall that as the circular cross section of a tube becomes elliptic in the passive scenario, the cross sectional area of the elliptic cross section becomes smaller than that of the circular cross section, and therefore a smaller volume of fluid will be accelerated and decelerated in the tube of elliptic cross section than in the circular one. It follows that the maximum flow rate reached at the peak of each oscillatory cycle might actually be higher in the tube of elliptic cross section. This was shown, however, \textsl{not} to be the case for steady flow in Figure~\ref{fig:steadyflowrate}. The reason for this is that the acceleration and deceleration peaks depend not only on the volume of fluid being accelerated but also on the opposition to that acceleration by the level and distribution of shear stress on the boundary.
However, even for steady flow, shear stress at the tube wall can be higher in the tube of elliptic cross section, and is in both scenarios for $\fa \ge 0.6$, as shown in Figure~\ref{fig:steadymaximumshearstress}. For pulsatile flow, Figure~\ref{fig:ActiveVsPassivemaxwallshearStressc1p0} shows that this remains true for pulsatile flow.

\begin{figure}[t]
     \centering
     \begin{subfigure}[b]{0.4\textwidth}
         \centering
         \includegraphics[width=\textwidth]{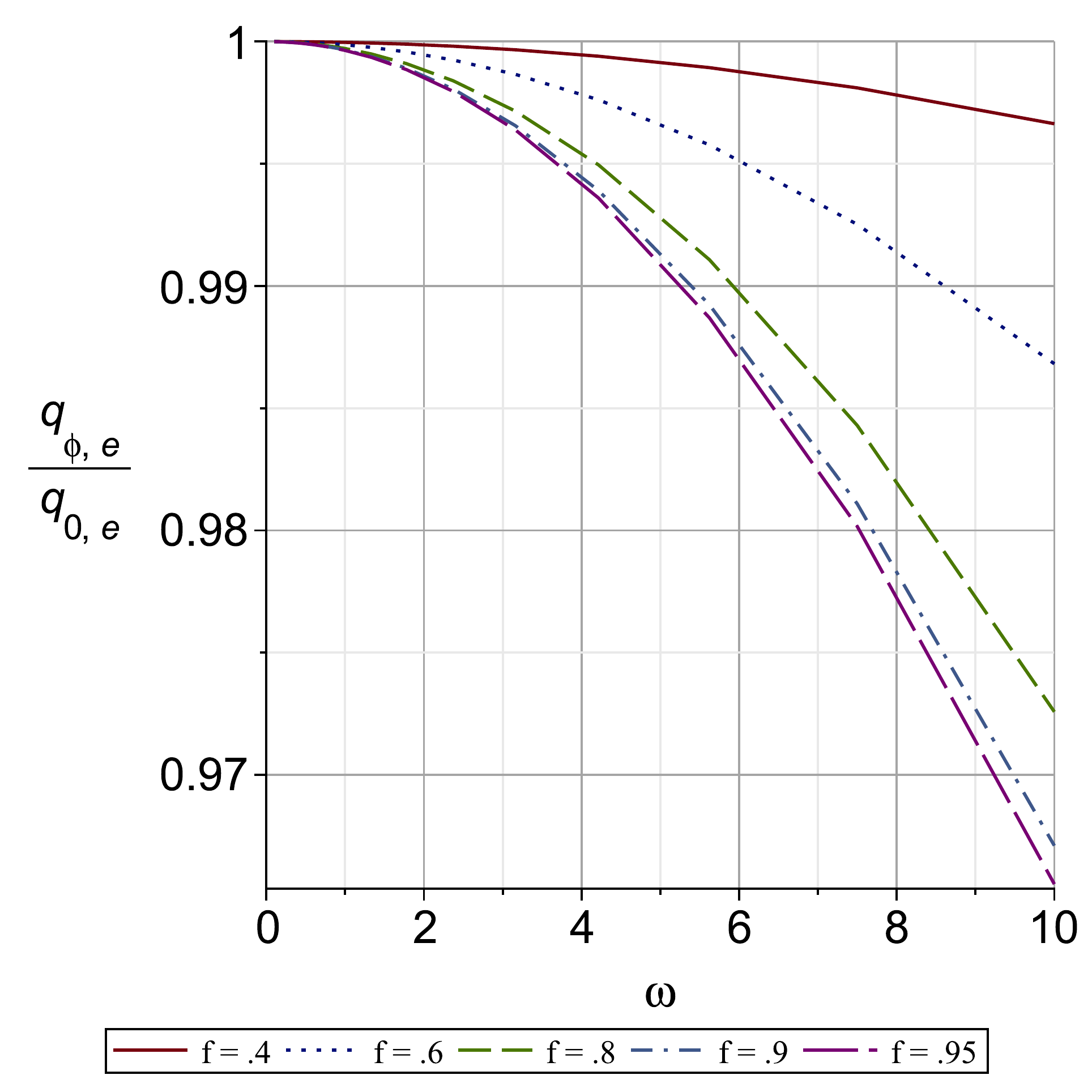}
         \caption{Passive, $c = 0.5$cm}
         \label{fig:Passivemaxflowratec0p5}
     \end{subfigure}
     \hfill
     \begin{subfigure}[b]{0.4\textwidth}
         \centering
         \includegraphics[width=\textwidth]{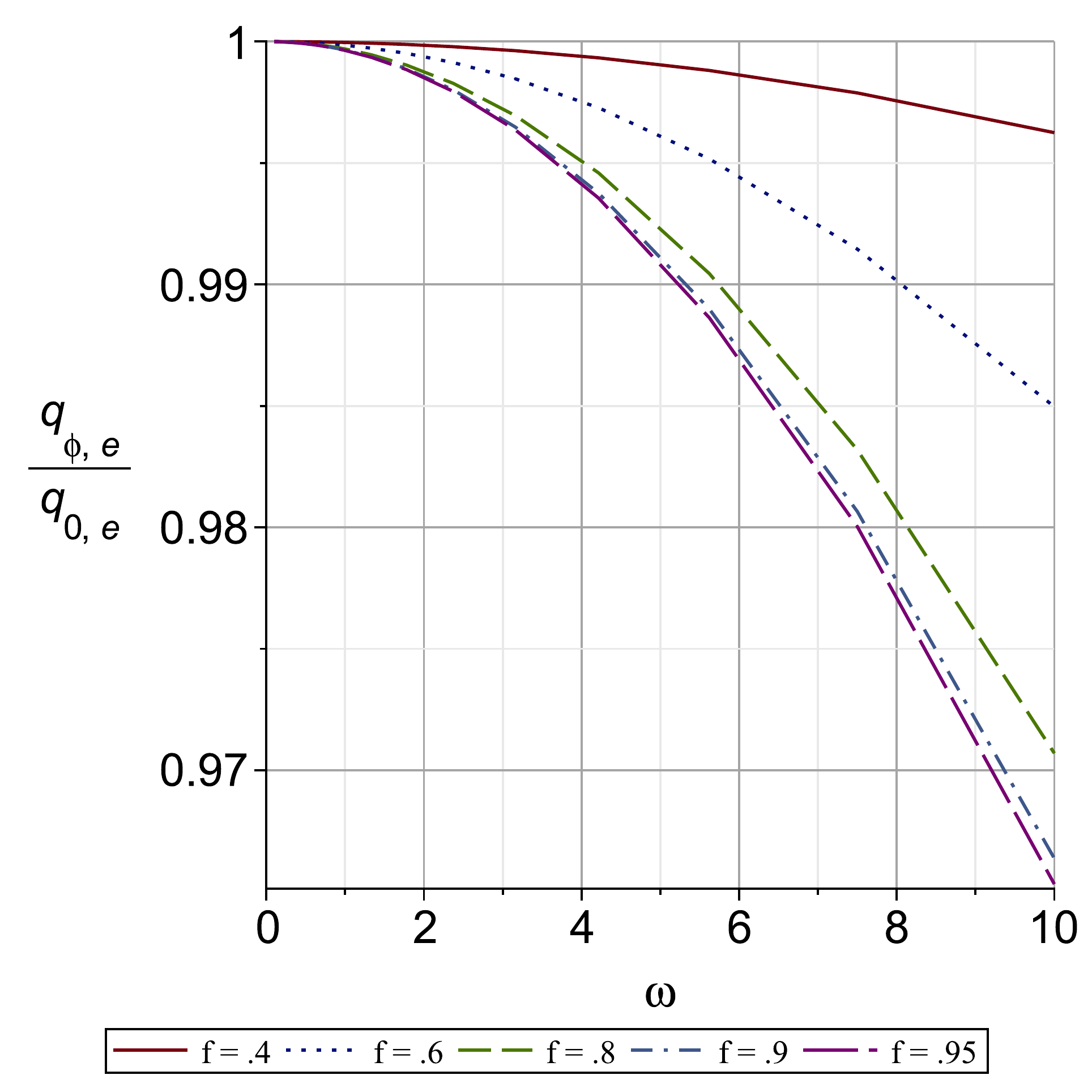}
         \caption{Active}
         \label{fig:Activemaxflowratec0p5}
     \end{subfigure}
        \caption{Oscillatory flow rate from equation~\eqref{eq:oscillatoryVolumetricFlow} compared in the passive scenario (left) to the active (right). The initial circumference was $c=0.5$cm. The same range of fractions $\fa$ are graphed for each scenario, namely $0.4$, $0.6$, $0.8$, $0.9$, and $0.95$ over the same range of frequencies $\omega$.}
        \label{fig:ActiveVsPassiveflowratec0p5}
\end{figure}

\begin{figure}[t]
     \centering
     \begin{subfigure}[b]{0.4\textwidth}
         \centering
         \includegraphics[width=\textwidth]{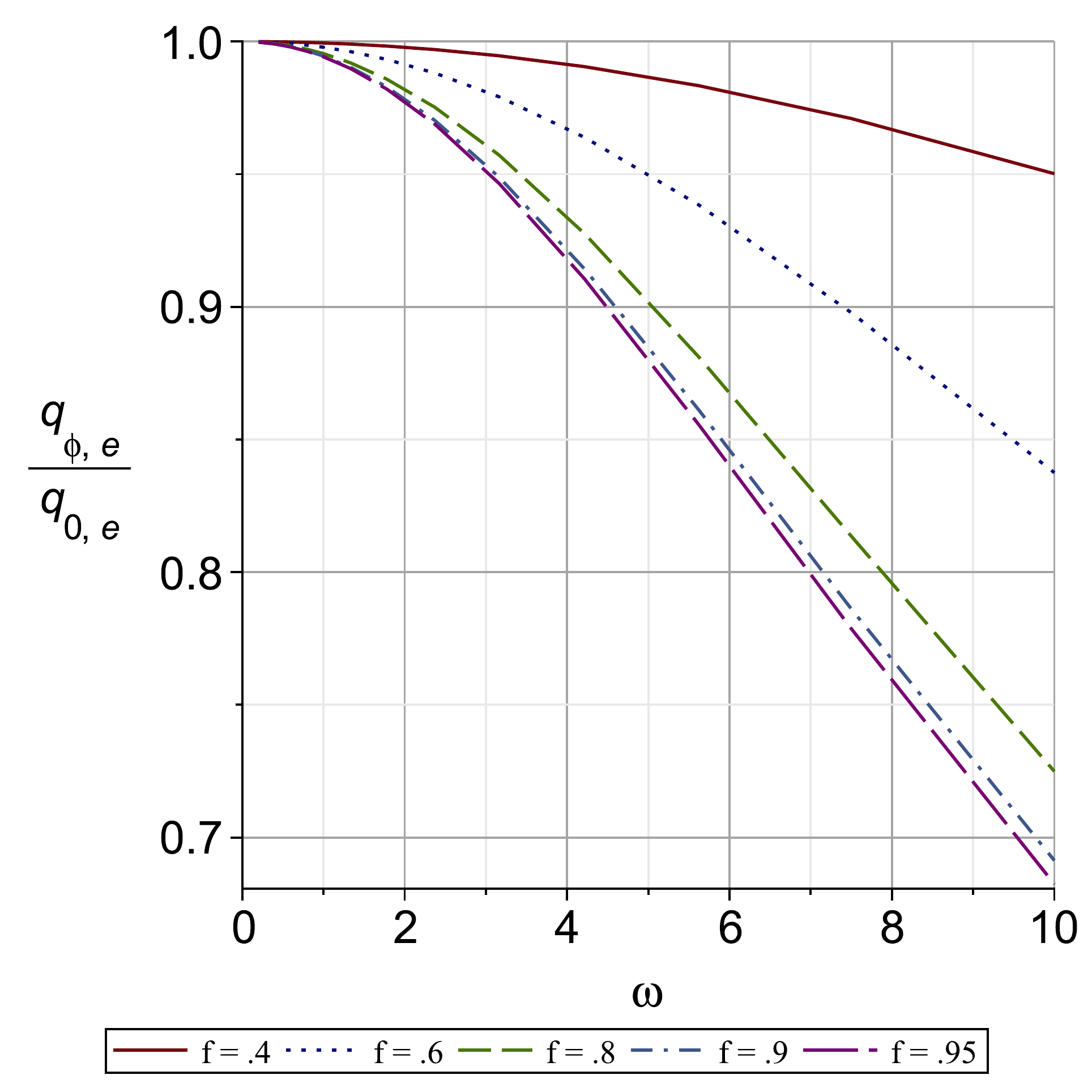}
         \caption{Passive, $c = 1.0$cm}
         \label{fig:Passivemaxflowratec1p0}
     \end{subfigure}
     \hfill
     \begin{subfigure}[b]{0.4\textwidth}
         \centering
         \includegraphics[width=\textwidth]{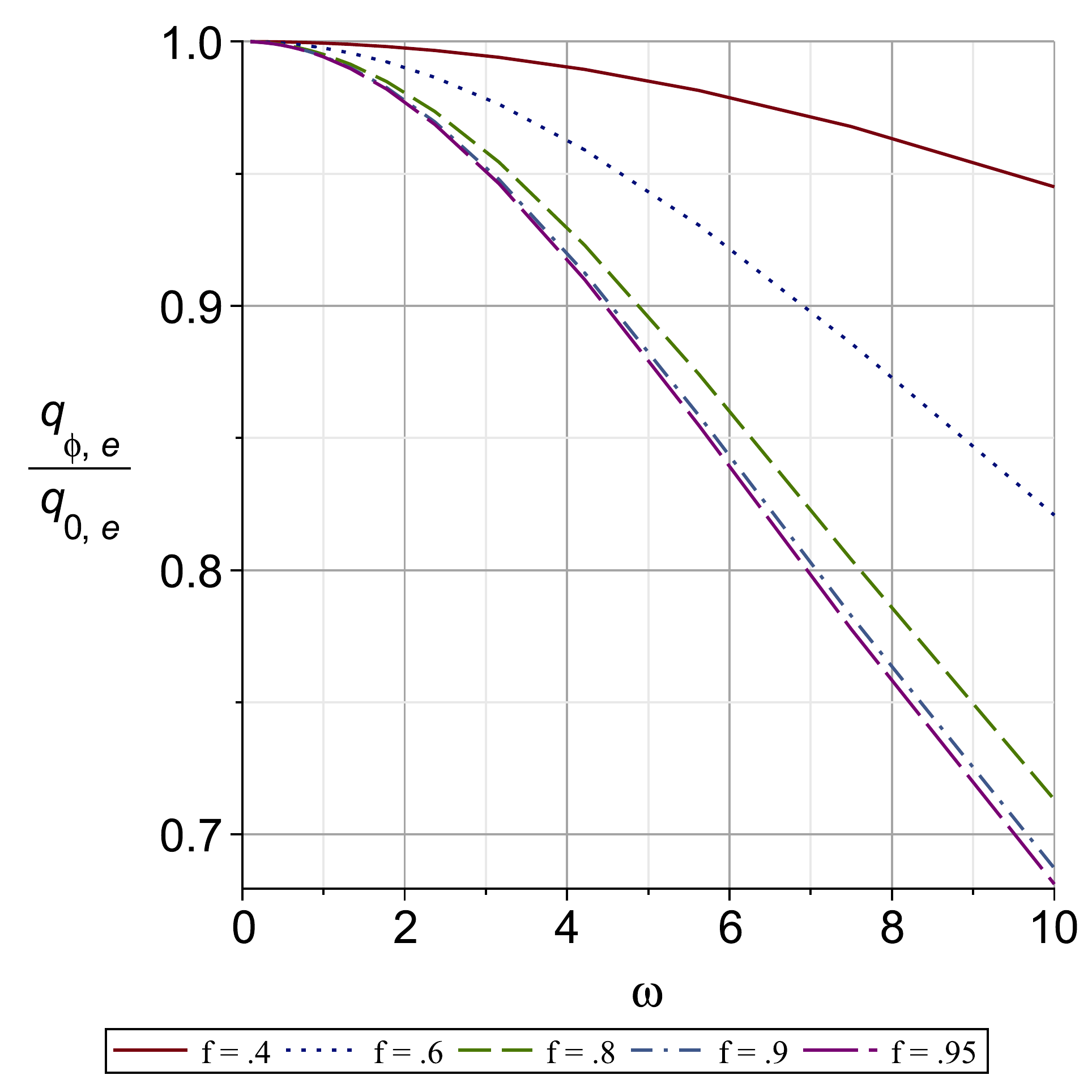}
         \caption{Active}
         \label{fig:Activemaxflowratec1p0}
     \end{subfigure}
        \caption{Oscillatory flow rate from equation~\eqref{eq:oscillatoryVolumetricFlow} compared in the passive scenario (left) to the active (right). The initial circumference was $c=1.0$cm. The same range of fractions $\fa$ are graphed for each scenario, namely $0.4$, $0.6$, $0.8$, $0.9$, and $0.95$ over the same range of frequencies $\omega$. These graphs do not at first look greatly different from those at $c=0.5$cm in Figure~\ref{fig:ActiveVsPassiveflowratec0p5}, but note the vertical scale: over this range of frequencies $\omega$ the flow rate drops by about $30$\% whereas with $c=0.5$ the flow rate dropped only by about $3$\%.}
        \label{fig:ActiveVsPassiveflowratec1p0}
\end{figure}

\begin{figure}[t]
     \centering
     \begin{subfigure}[b]{0.4\textwidth}
         \centering
         \includegraphics[width=\textwidth]{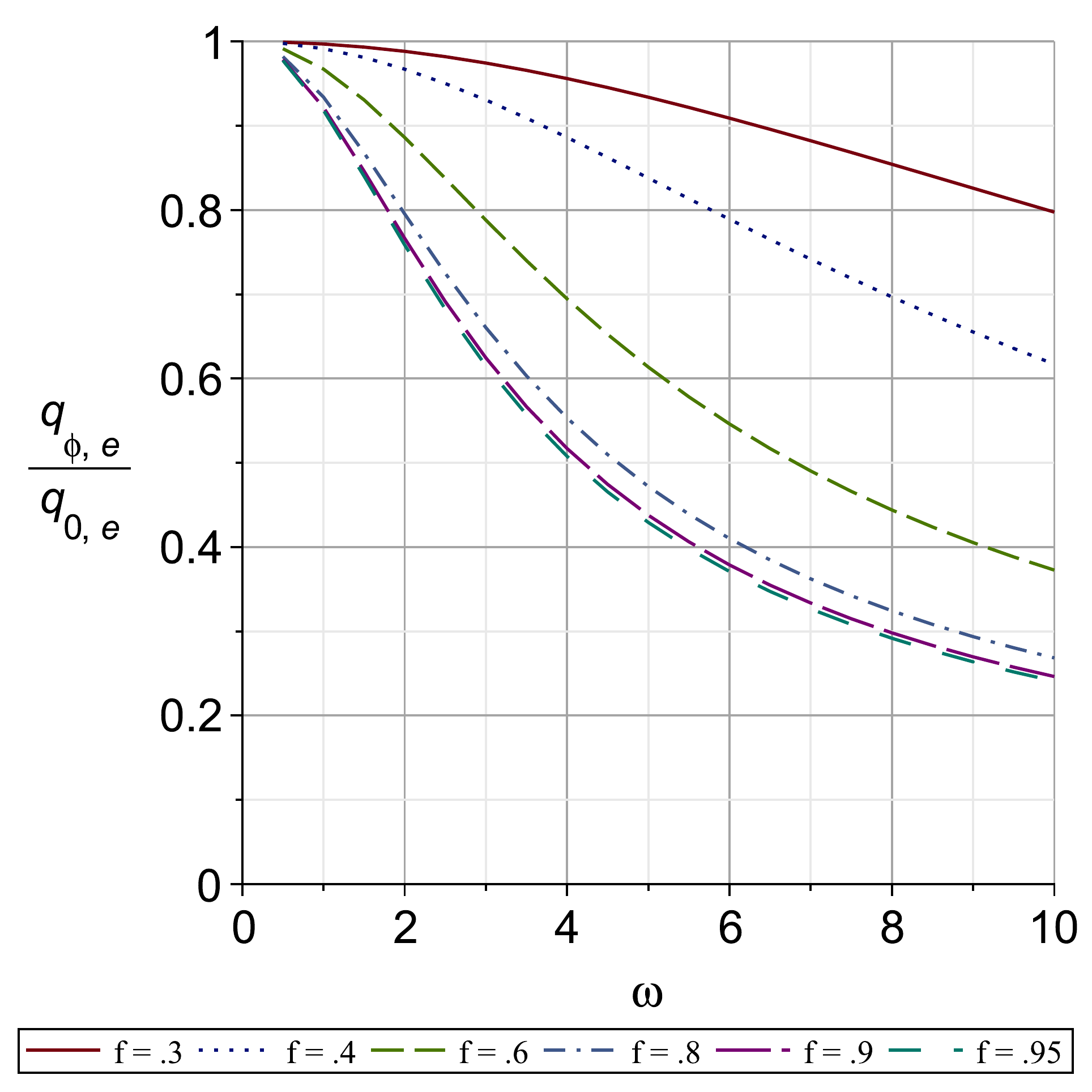}
         \caption{Passive, $c = 2.0$cm}
         \label{fig:Passivemaxflowratec2p0}
     \end{subfigure}
     \hfill
     \begin{subfigure}[b]{0.4\textwidth}
         \centering
         \includegraphics[width=\textwidth]{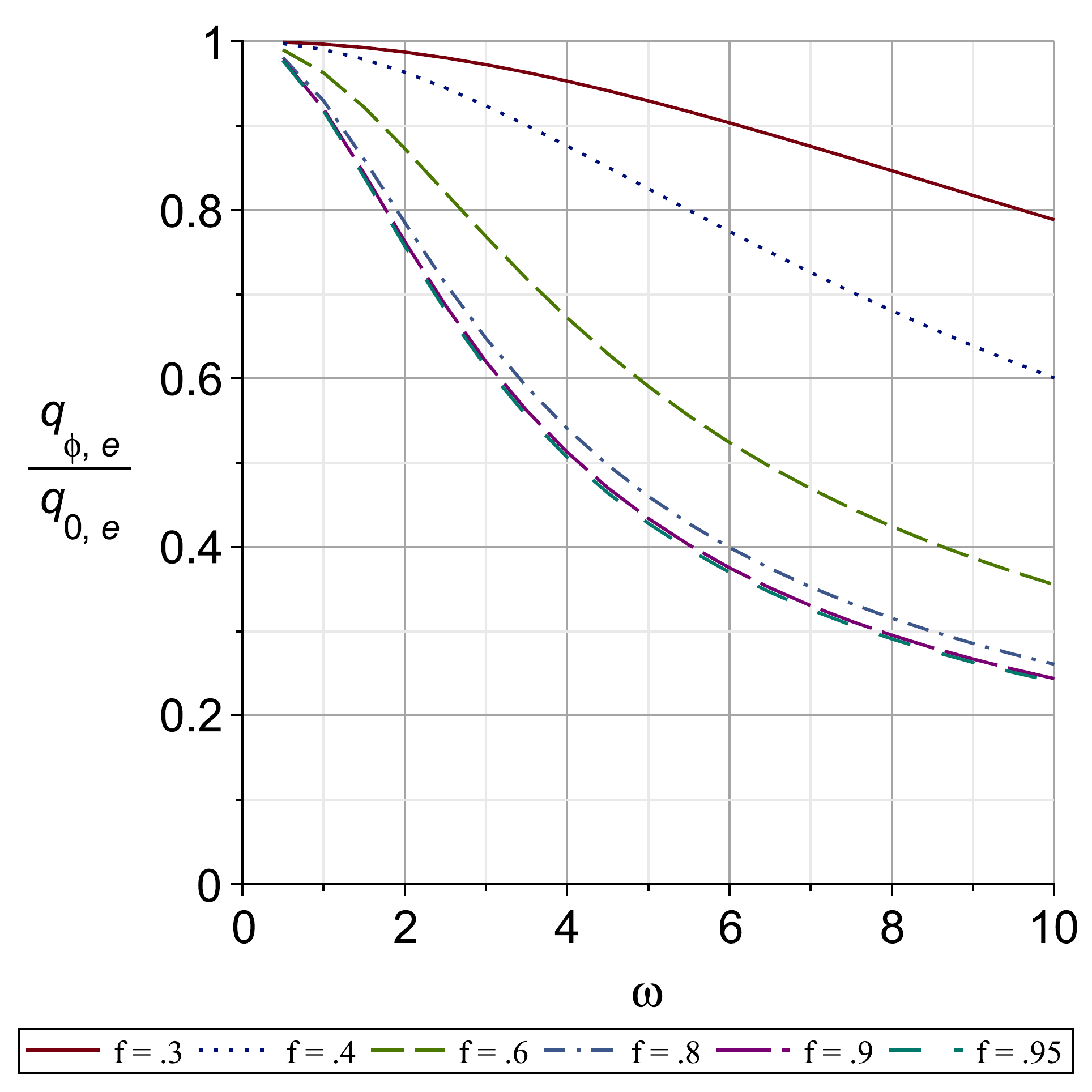}
         \caption{Active}
         \label{fig:Activemaxflowratec2p0}
     \end{subfigure}
        \caption{Oscillatory flow rate from equation~\eqref{eq:oscillatoryVolumetricFlow} compared in the passive scenario (left) to the active (right). The initial circumference was $c=2.0$cm. The same range of fractions $\fa$ are graphed for each scenario, namely $0.3$, $0.4$, $0.6$, $0.8$, $0.9$, and $0.95$ over the same range of frequencies $\omega$. These graphs show a much greater dependence on $\omega$ than those of either $c=1.0$cm or $c=0.5$cm in figures~\ref{fig:ActiveVsPassiveflowratec1p0} and~\ref{fig:ActiveVsPassiveflowratec0p5}.}
        \label{fig:ActiveVsPassiveflowratec2p0}
\end{figure}

The pulsatile flow rates shown in Figures~\ref{fig:ActiveVsPassiveflowratec0p5}--\ref{fig:ActiveVsPassiveflowratec2p0} do not become higher for tubes of elliptic cross section than in circular cross section, even in the active scenario. They show moderate dependence on imposed frequency $\omega$ with a general downward trend with increasing frequency.  They also show a decrease in maximum flow for tubes with smaller fractions $\fa$ of the original radius.  They do not show a great effect of the two different scenarios, active versus passive.  Indeed, those figures show that there is very little effect on maximum flow rate between the scenarios.  The maximum velocity profiles show similar behaviour, and are not plotted here as being redundant.

Tubes of elliptic cross section with smaller $\fa$ show greater dependence of maximum pulsatile flow rate (and similarly velocity, not shown) on the imposed frequency.  Conversely, tubes of smaller $\fa$ show weaker dependence of shear stress on the imposed frequency $\omega$.

As a final consideration, the dependence of oscillatory properties on frequency $\omega$ seen in the figures can be interpreted as the way the first few harmonics of a composite pressure wave would be individually affected under the two scenarios being considered.  We do not otherwise pursue here the idea of a composite pressure wave.

\begin{figure}[t]
     \centering
     \begin{subfigure}[b]{0.4\textwidth}
         \centering
         \includegraphics[width=\textwidth]{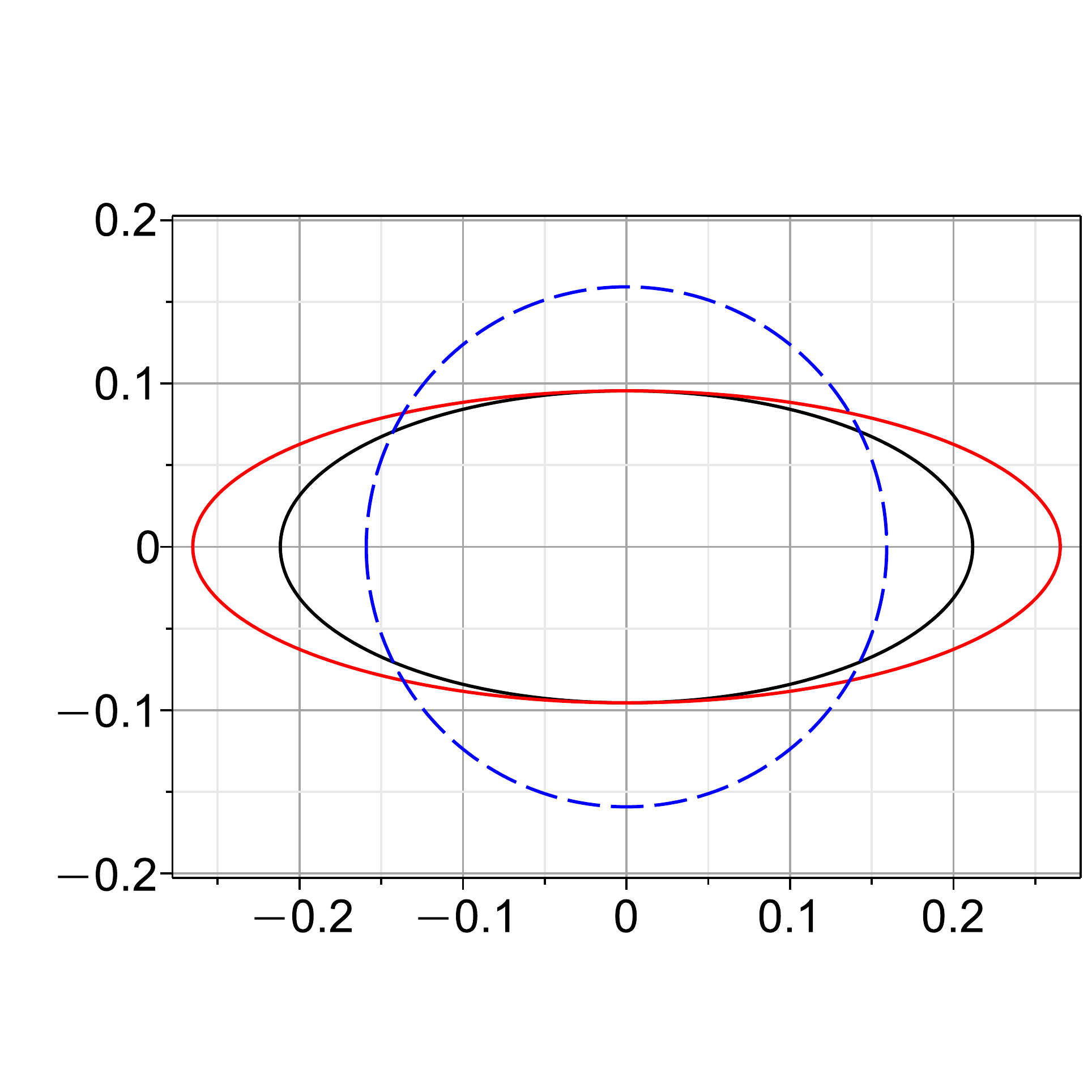}
         \caption{Active (red) vs Passive (black) $\fa = 0.6$}
         \label{fig:ActiveVsPassiveEllipses}
     \end{subfigure}
     \hfill
     \begin{subfigure}[b]{0.4\textwidth}
         \centering
         \includegraphics[width=\textwidth]{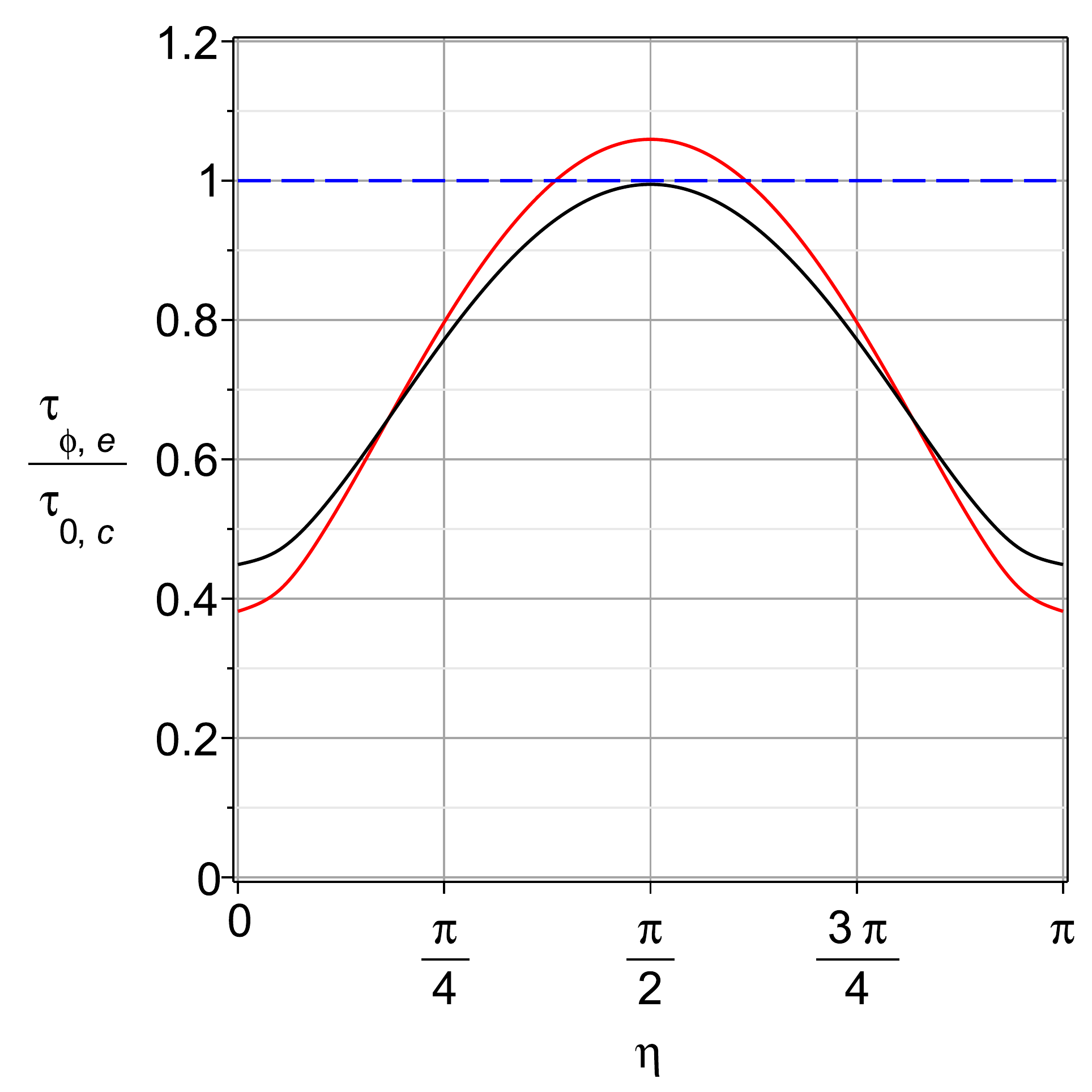}
         \caption{Peak shear stress $\fa = 0.6$}
         \label{fig:WallShearStress}
     \end{subfigure}
        \caption{Left: A blood vessel of circular cross section with circumference $1.0$cm and radius $a=1.0/(2\pi)$cm (blue dashed line) is deformed by imposed vertical forces to ellipses with semiminor axis $\beta = \fa a$ with $\fa=0.6$, in two different ways: under the active scenario, neuromuscular control relaxes the vessel wall so it stretches in order to maintain the initial cross-sectional area (red curves), and under the passive scenario, the vessel wall stays at $1.0$cm in circumference (black curves), thereby losing some cross-sectional area.  Right: The resulting peak shear stress over one cycle of oscillation from equation~\eqref{eq:oscillatoryShearStress}, scaled by the steady wall stress $\hat \tau_{0,c}$ from equation~\eqref{eq:steadMaximumShearStress} in the case $\alpha=\beta=a$, the radius of the circle. Both scenarios are graphed at their peaks in time, together with the wall stress from equation~\eqref{eq:circlewallshearstress} (blue dashed line) from the original circle. The wider elliptical vessel has the smallest minimum oscillatory wall shear stress, although the maximum is greater than that of the original circle.}
        \label{fig:ActiveVsPassive}
\end{figure}

\begin{figure}[t]
     \centering
     \begin{subfigure}[b]{0.4\textwidth}
         \centering
         \includegraphics[width=\textwidth]{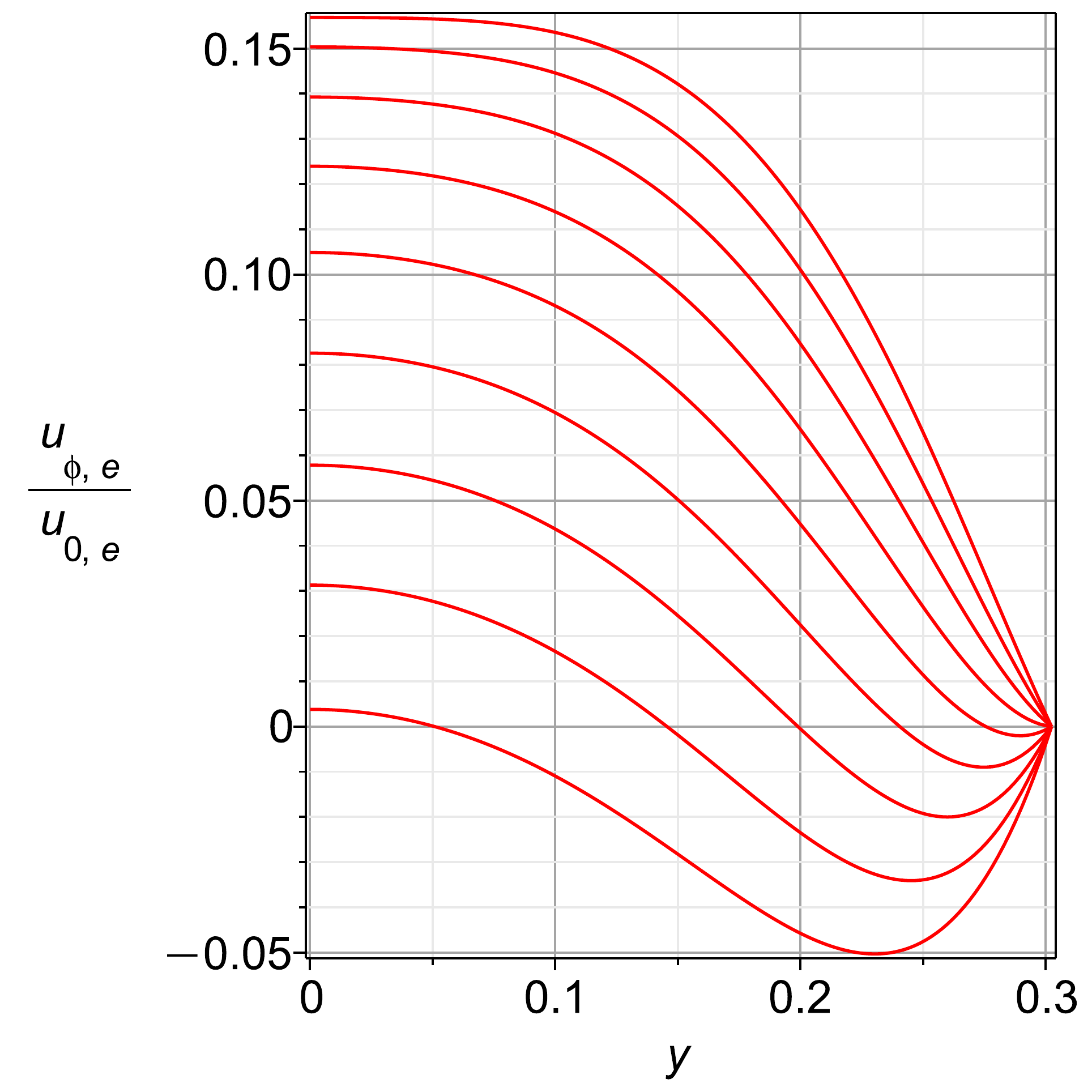}
         \caption{minor axis profile $0 \le y \le \beta$}
         \label{fig:quartercycleprofile}
     \end{subfigure}
     \hfill
     \begin{subfigure}[b]{0.4\textwidth}
         \centering
         \includegraphics[width=\textwidth]{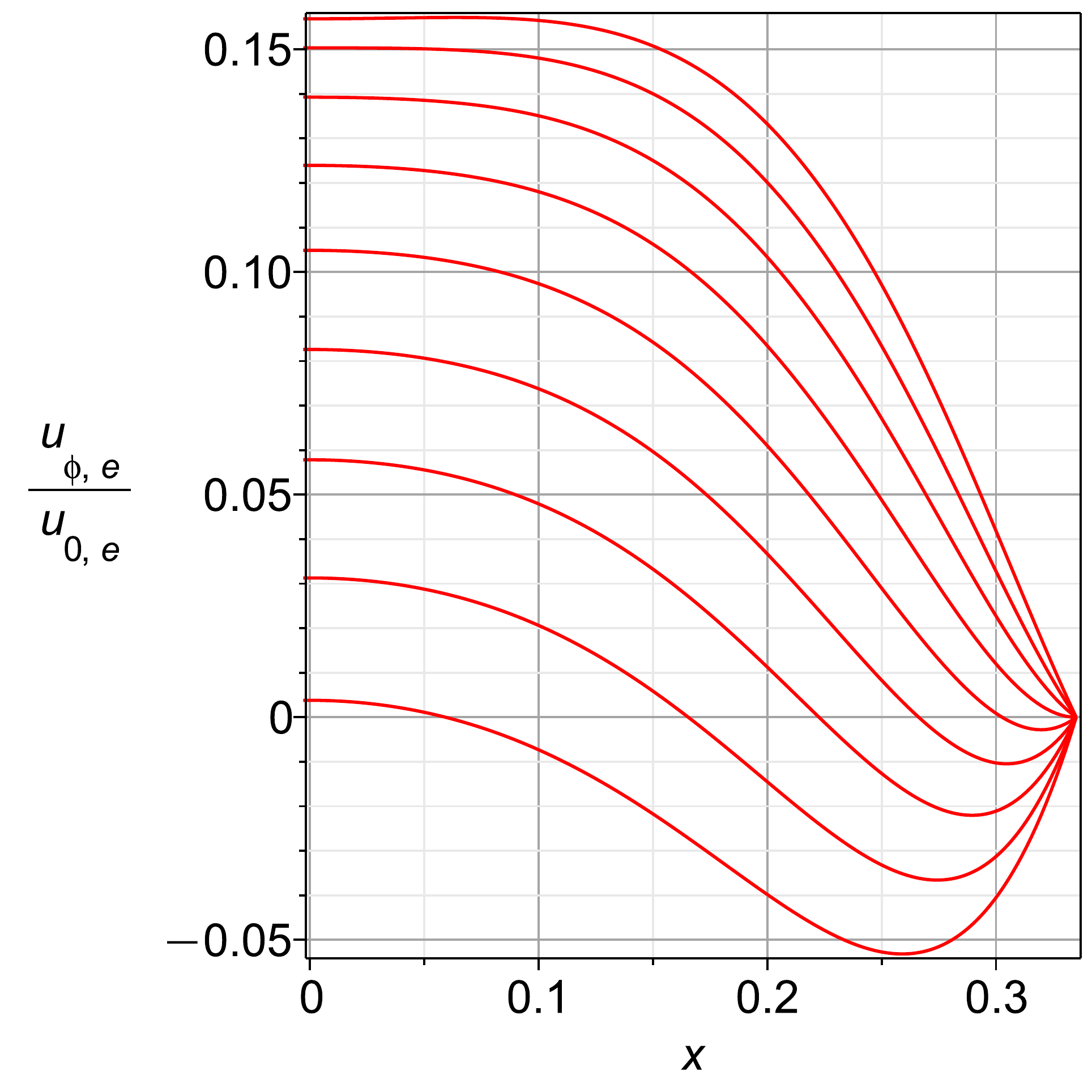}
         \caption{major axis profile $0 \le x \le \alpha$}
         \label{fig:quartercyclesection}
     \end{subfigure}
        \caption{One quarter of a cycle of an oscillation starting at a maximum, with each curve showing (half) the velocity profile from equation~\eqref{eq:velocityProfiles} at successive instants.  The top curve is at time $t=0$. As the time $t$ is sampled over the quarter cycle, the curves go \textsl{down}. In the next quarter cycle (not shown) they would go further down to the minimum; then they would come up over the next half-cycle (not shown, because the overlapping lines would be harder to interpret) to the beginning again. The particulars of this picture are that the original circumference was $c=2.0$cm, the semiminor axis of the ellipse is a fraction $\fa = 0.95$ of the original circular radius, the scenario is active so that the semimajor axis is $1/\fa = 1.053$ of the original circular radius, and the frequency is $\omega = 11.2847265810330$Hz (which is slightly higher than for most of our simulations). This is, as it happens, at a double eigenvalue, which is why so many decimal places of the frequency are printed. Fractions $\fa$ very close to $1$ are harder for the Mathieu expansion to model, as indicated in the text, but this fraction, $\fa = 0.95$, is routine, and slight deformations from circular cross section are more likely to occur, so this is a reasonably realistic case.}
        \label{fig:Velocityprofile}
\end{figure}
\begin{figure}[t]
     \centering
     \begin{subfigure}[b]{0.4\textwidth}
         \centering
         \includegraphics[width=\textwidth]{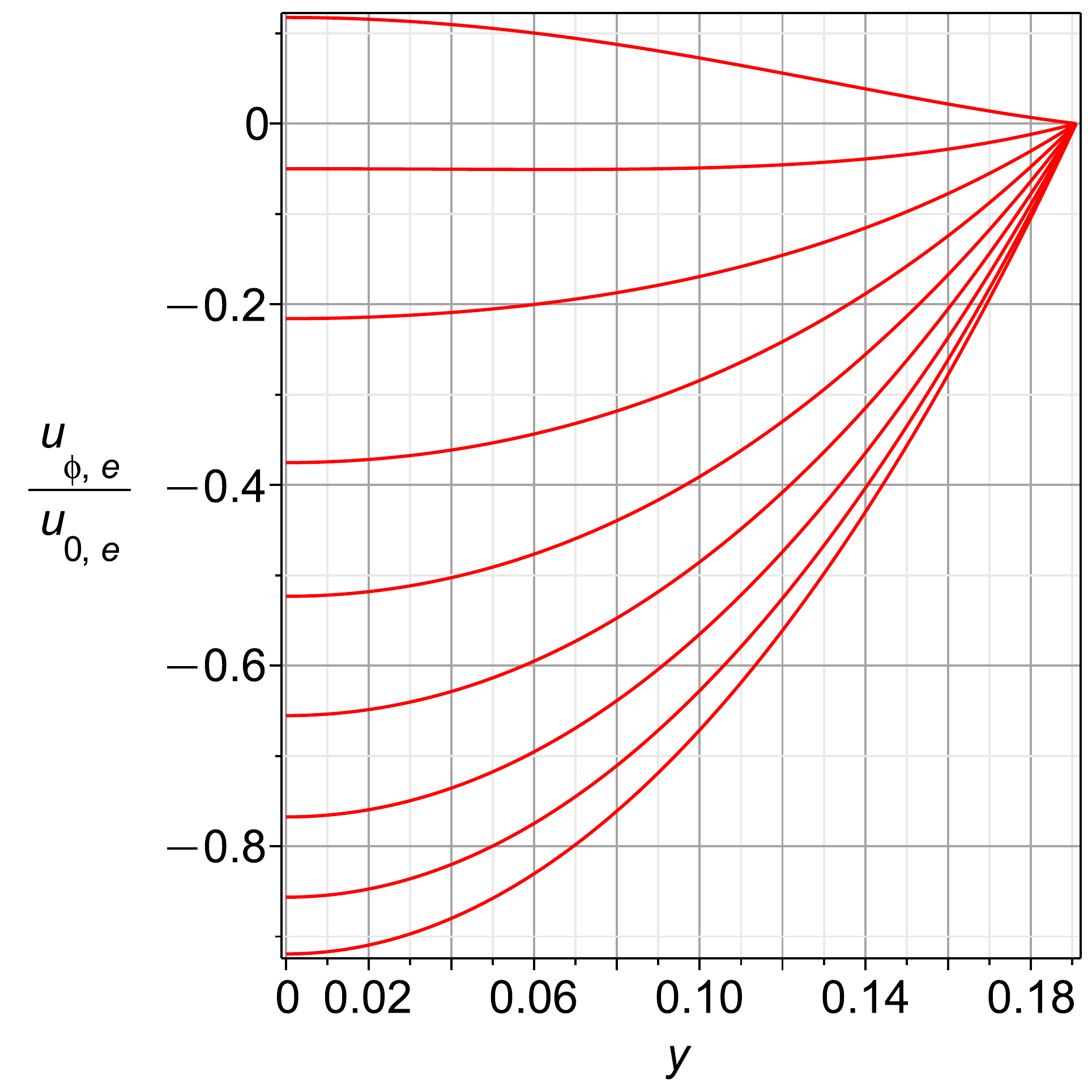}
         \caption{minor axis profile $0 \le y \le \beta$}
         \label{fig:quartercycleprofilef6}
     \end{subfigure}
     \hfill
     \begin{subfigure}[b]{0.4\textwidth}
         \centering
         \includegraphics[width=\textwidth]{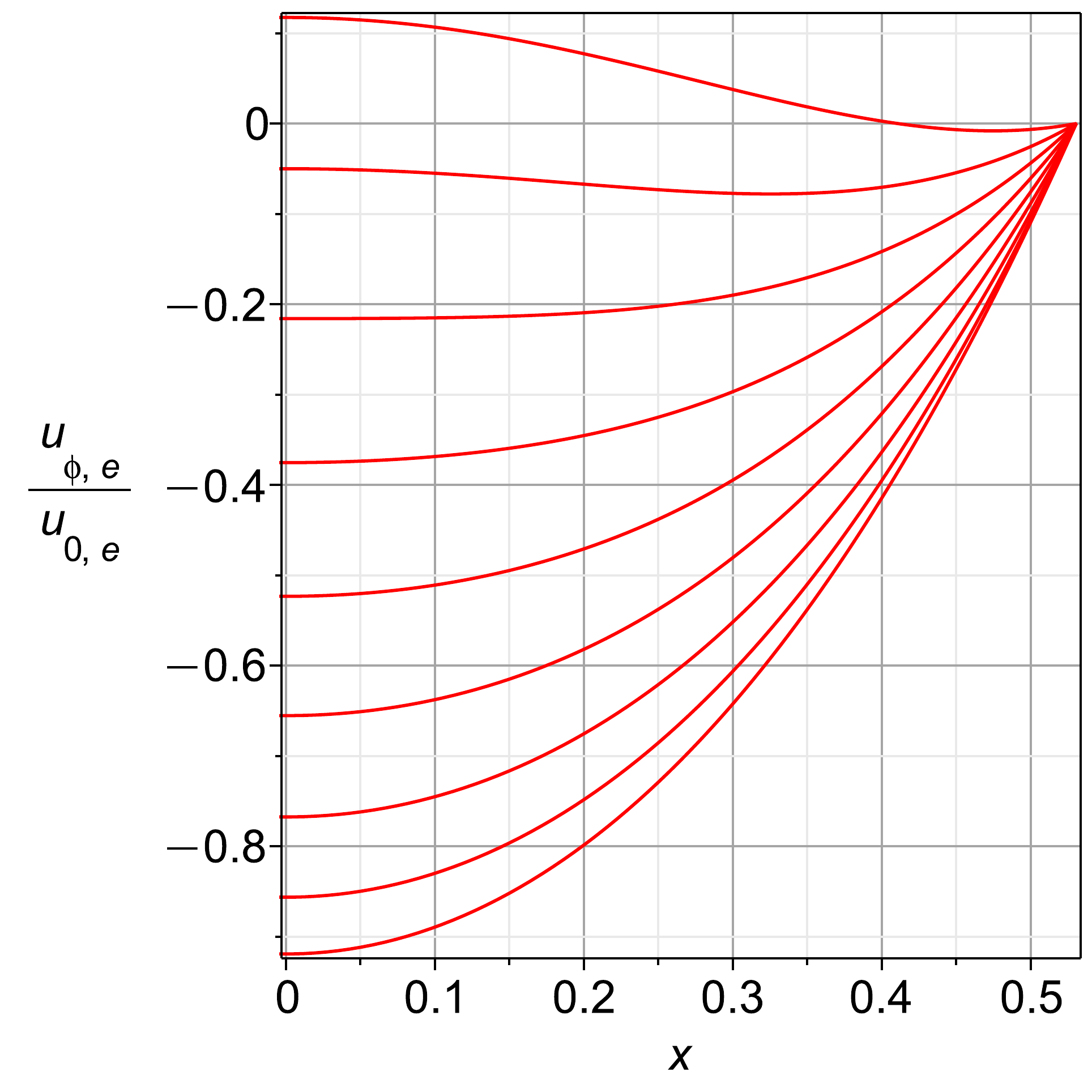}
         \caption{major axis profile $0 \le x \le \alpha$}
         \label{fig:quartercyclesectionf6}
     \end{subfigure}
        \caption{One quarter of a cycle of an oscillation, starting at time $t=0$ half-way between the minimum and the maximum. Each curve shows (half) the velocity profile from equation~\eqref{eq:velocityProfiles} at successive instants.  Successive curves show the profile going \textsl{down} as the sampled times increase. In the next quarter cycle (not shown) they would go \textsl{up}; then they would go up further to the maximum for the cycle, then come back down over a final quarter cycle to the beginning again (not shown because the overlapping curves would be harder to interpret). The particulars are that the original circular circumference was $c=2.0$cm, the semiminor axis is a fraction $\fa = 0.6$ of the original radius, the scenario is an active one and so the semimajor axis is $1/\fa = 1.667$ of the original radius, and the frequency $\omega = 0.959305048332072$Hz which is well within the frequency range of most of our simulations. This is again, as it happens, at a double eigenvalue, which is why so many decimal places of the frequency are printed.  The double eigenvalue solution is very similar to nearby simple eigenvalue solutions, as discussed in the text.  The fraction $\fa = 0.6$ seems reasonable for the active scenario, because according to figure~\ref{fig:ConstantAreaCircumferenceRatio} the circumference would need to stretch by perhaps $25$\%. } 
        \label{fig:Velocityprofilef6}
\end{figure}

\begin{figure}[t]
     \centering
     \begin{subfigure}[b]{0.48\textwidth}
         \centering
         \includegraphics[width=\textwidth]{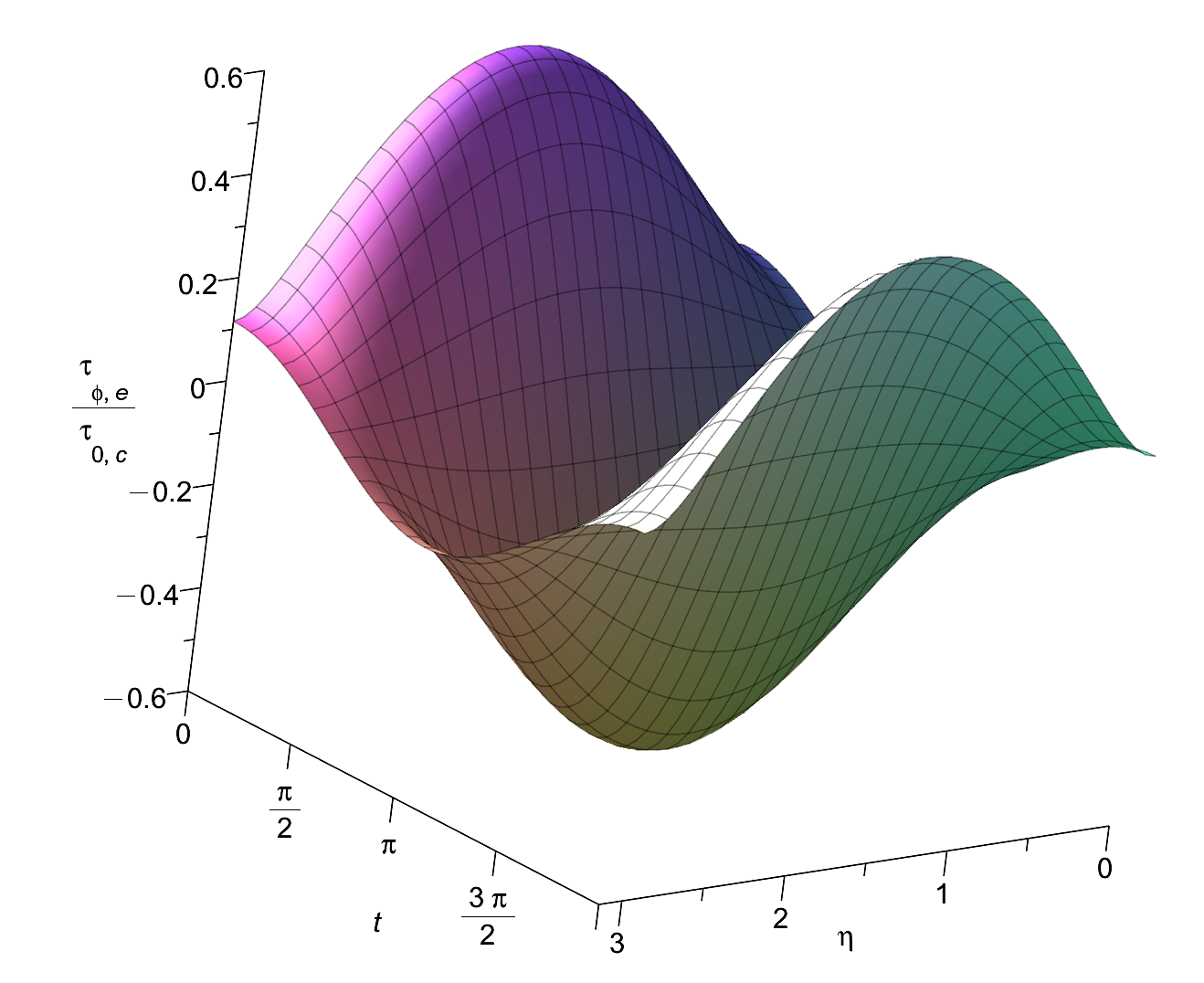}
         \caption{Passive, $c = 1.0$cm}
         \label{fig:PassiveShearStress3D}
     \end{subfigure}
     \hfill
     \begin{subfigure}[b]{0.4\textwidth}
         \centering
         \includegraphics[width=\textwidth]{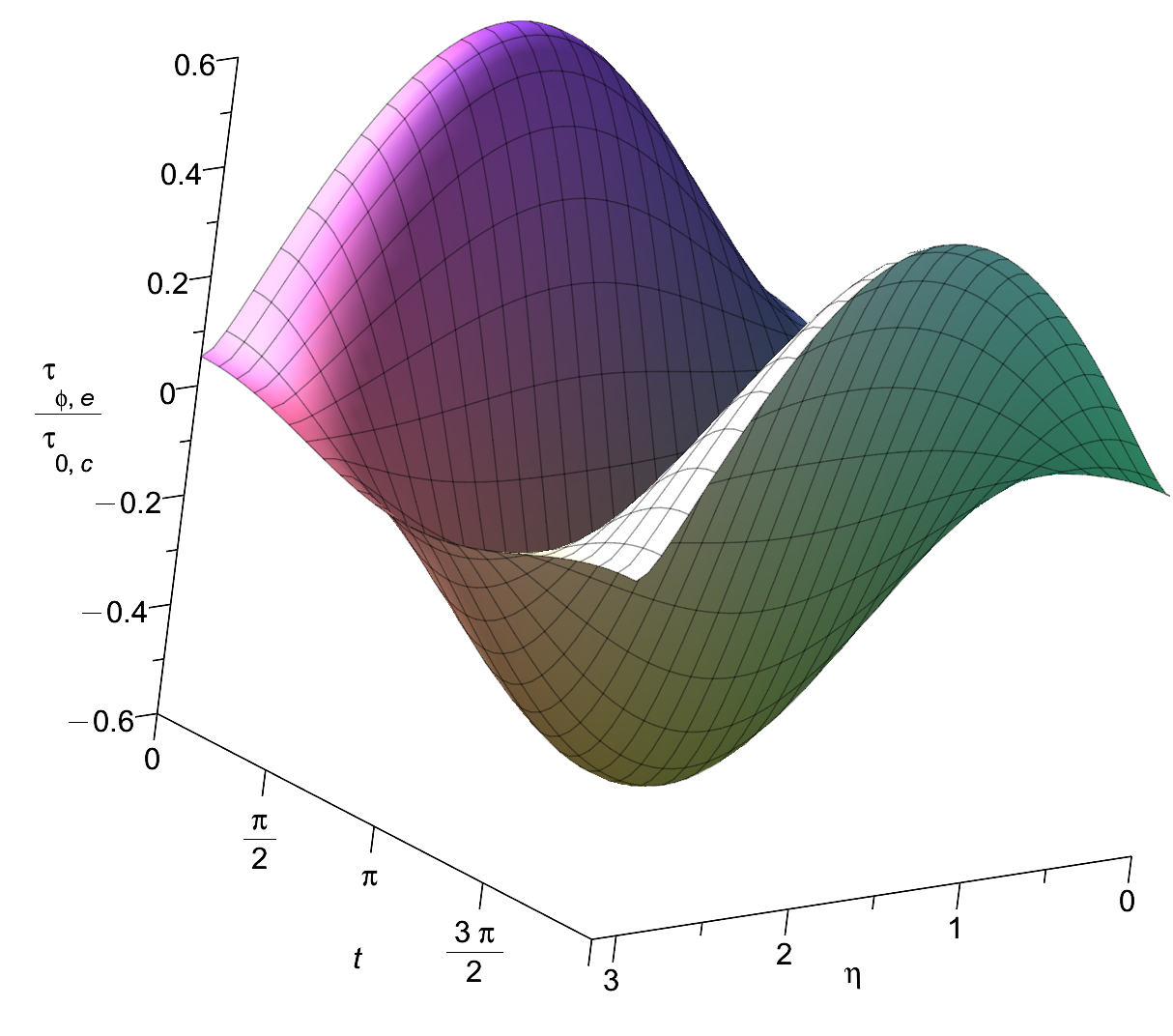}
         \caption{Active}
         \label{fig:ActiveShearStress3d}
     \end{subfigure}
        \caption{The real part of the oscillatory wall shear stress $\tau_e/\hat{\tau}_{0,e}$ throughout a cycle from equation~\eqref{eq:oscillatoryShearStress} compared in the passive scenario (left) to the active (right). The initial circumference was $c=2.0$cm. The fraction of the original circular radius is $\fa=0.3$ in both cases, and in both cases $\omega=1$ Hz. The main difference between the two figures, which is slight, occurs at the poles of the major axis, where the shear stress is least.  There the wall shear stress is smaller in the active scenario than it is in the passive scenario.}
        \label{fig:ActiveVsPassiveShearStress3d}
\end{figure}

\begin{figure}[t]
     \centering
     \begin{subfigure}[b]{0.4\textwidth}
         \centering
         \includegraphics[width=\textwidth]{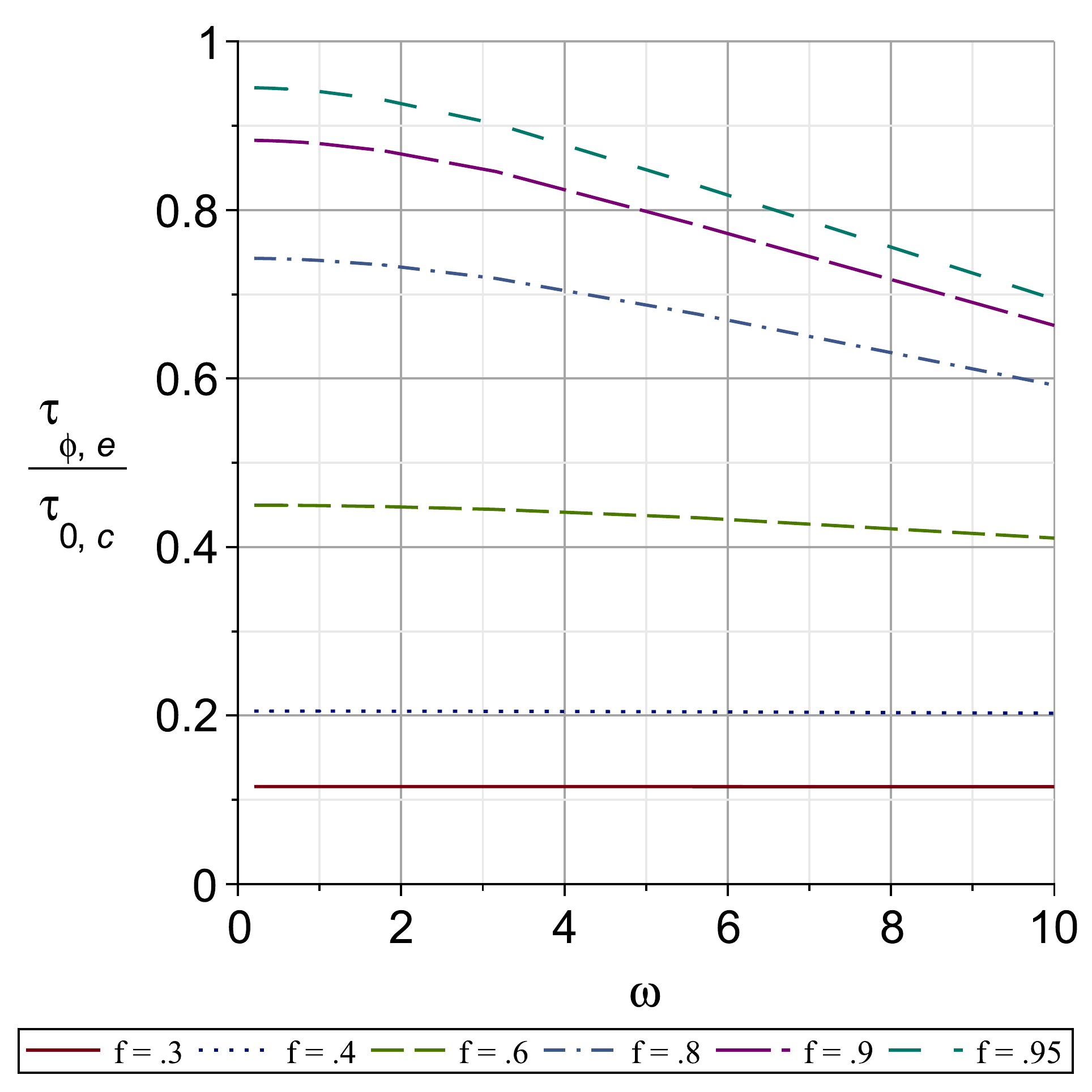}
         \caption{Passive, $c = 1.0$cm}
         \label{fig:PassiveWallShearStressc1p0}
     \end{subfigure}
     \hfill
     \begin{subfigure}[b]{0.4\textwidth}
         \centering
         \includegraphics[width=\textwidth]{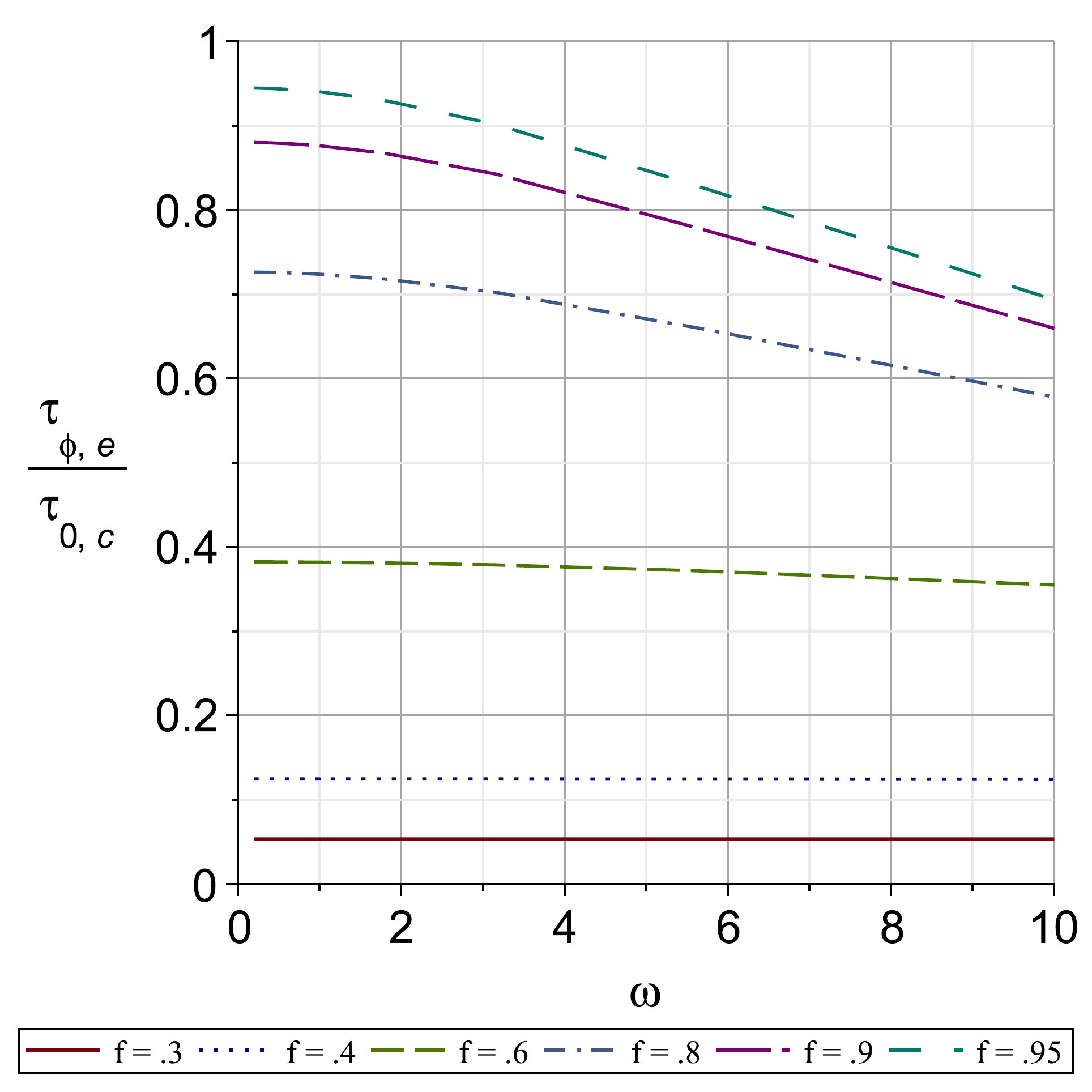}
         \caption{Active}
         \label{fig:ActiveWallShearStressc1p0}
     \end{subfigure}
        \caption{Minimum oscillatory wall shear stress from equation~\eqref{eq:oscillatoryShearStress} compared in the passive scenario (left) to the active (right). The initial circumference was $c=1.0$cm. The same range of fractions $\fa$ are graphed for each scenario, namely $0.3$, $0.4$, $0.6$, $0.8$, $0.9$, and $0.95$ over the same range of frequencies $\omega$. In this figure we see that the different scenarios do not produce greatly different minimum wall shear stresses, with the greatest differences appearing for the smallest fractions $\fa$; and moreover that the different scenarios show a similar dependence on the imposed frequency $\omega$, with a general decrease in shear stress for higher frequency. This dependence is weaker with the more compressed cross sections (smaller fractions $\fa$).  We used the simple symbol $f$ instead of $\fa$ in the legend to improve readability.}
        \label{fig:ActiveVsPassiveWallShearStressc1p0}
\end{figure}


\section{Concluding remarks}
The problem of pulsatile flow in tubes of elliptic cross sections is important from a physiological as well as mathematical perspective, and the aim of our study was to examine this problem from both of these perspectives, using a tube of elliptic cross section as a model of a deformed blood vessel. While this is clearly a simplified model of the many ways in which a blood vessel may be deformed, it allowed us to explore a full range of distortions of a tube of circular cross section, from being fully open to almost closed.

More important than the final form into which a vessel is deformed are the constraints and scenarios under which the transformation from circular to elliptic cross section takes place. The two scenarios which we have considered highlight the mathematical and physiological aspects of the problem and provide useful information on the way the neurovascular control system may respond to the deformation of a blood vessel in the physiological setting. In particular, the ability of the control system to maintain a constant cross sectional area under the active scenario is clearly limited to only small or moderate departures from the circular cross section. When the departure from circular cross section is large (high ellipticity, low fraction $\fa$), a prohibitively large increase in the circumference of the wall would be required to maintain the cross sectional area available for the flow, as illustrated in Figure~\ref{fig:ConstantAreaCircumferenceRatio}.

We have extended both the scope and the range of results currently available for this problem by using new methodology to overcome difficulties encountered in the solution of the governing Mathieu equations and in the numerical evaluation of Mathieu functions in the past. Specifically, we used a careful spectral method, including explicit solution in the case of double eigenvalues, for the solution of the governing equations. We used extended precision where necessary to overcome issues of ill-conditioning, for $\fa$ very close to $1$ which is paradoxically the difficult case. We believe that this novel approach offers a useful new tool in further study of pulsatile blood flow under various pathological conditions.

\bibliographystyle{plain}

\end{document}